\documentclass[traditabstract]{aa}
\usepackage{graphicx}
\usepackage{amsmath,amssymb}
\usepackage{natbib}
\bibpunct{(}{)}{;}{a}{}{,}

\newcommand{\grad}{\vec \nabla}

\newcommand{\MUSIC}{{\sf MUSIC}}

\begin{document}
   \title{A Jacobian-free Newton-Krylov method for time-implicit multidimensional hydrodynamics}
   \subtitle{Physics-based preconditioning for sound waves and thermal diffusion}
  \author{M. Viallet\inst{\ref{inst1}}
    \and T. Goffrey\inst{\ref{inst2}}
    \and I. Baraffe\inst{\ref{inst2},\ref{inst3},\ref{inst1}}
    \and D. Folini\inst{\ref{inst3}}
    \and C. Geroux\inst{\ref{inst2}}
    \and M. V. Popov\inst{\ref{inst3}}
    \and J. Pratt\inst{\ref{inst2}}
    \and R.Walder\inst{\ref{inst3}}}

   \institute{Max$-$Planck$-$Institut f\"ur Astrophysik, Karl Schwarzschild Strasse 1, Garching, D-85741, Germany\\
   \email{mviallet@mpa-garching.mpg.de}\label{inst1}
   \and College of Engineering, Mathematics and Physical Sciences, University of Exeter, Exeter, EX4 4QL, UK \label{inst2}
   \and Ecole Normale Sup\'erieure, Lyon, CRAL, UMR CNRS 5574, Universit\'e de Lyon, France \label{inst3}}

   \date{Received; accepted}

  \abstract{This work is a continuation of our efforts to develop an efficient implicit solver for multidimensional hydrodynamics for the purpose of studying important physical processes in stellar interiors, such as turbulent convection and overshooting. We present an implicit solver resulting from the combination of a Jacobian-Free Newton-Krylov method and a preconditioning technique tailored for the inviscid, compressible equations of stellar hydrodynamics. We assess the accuracy and performance of the solver for both 2D and 3D problems, for Mach numbers down to $10^{-6}$. Although our applications concern flows in stellar interiors, the method can be applied to general advection and/or diffusion dominated flows. The method presented in this paper opens up new avenues in 3D modeling of realistic stellar interiors allowing the study of important problems in stellar structure and  evolution.}


  \titlerunning{A JFNK method for time-implicit multi-D hydrodynamics.}
  \authorrunning{M.Viallet et al.}

 \maketitle

\section{Introduction}

The transport of heat, chemical species, and angular momentum in stellar interiors is governed by three dimensional, nonlinear (magneto-)hydrodynamical processes which develop over a wide range of temporal and spatial scales. The study of these processes with numerical simulations is a powerful way to improve our understanding of stellar structure and stellar evolution. Unfortunately, the integration of the compressible hydrodynamical equations with time explicit methods comes with a constraint on the time step resulting from the propagation of sound waves. This is the well-known Courant-Friedrich-Lewy (CFL) stability condition. We define CFL$_\mathrm{hydro}$ as the ratio between the time step and the largest explicit time step allowed by the CFL condition:

\begin{equation}
\mathrm{CFL}_\mathrm{hydro} = \max \frac{\left( |\vec{u}| + c_s\right )  \Delta t}{\Delta x},
\end{equation}

\noindent where $\Delta t$ is the time step, $\Delta x$ the mesh spacing, $c_s$ the adiabatic sound speed, and $\vec{u}$ the flow velocity. Time explicit methods require CFL$_\mathrm{hydro} \lesssim 1$. This results in values of the time step which are small compared to the typical time scale of the relevant processes (e.g., the convective turnover time scale), making this approach computationally demanding. Nevertheless, an explicit time integration method remains the method of choice for multi-dimensional hydrodynamics in the astrophysical community \citep[see e.g.][]{bazan_djehuty_2003,meakin_turbulent_2007, mocak_multidimensional_2011,herwig_global_non_spherical_2014}. A way to overcome this limitation is to rely on sound-proof models, which filter sound waves. Popular sound-proof models are the Boussinesq, the anelastic, or the pseudo-incompressible models \cite[see e.g.][for a review]{glatzmaier2013introduction}. The use of such models, however, comes at the cost of a restricted range of applications due to the underlying approximations. Ideally, one seeks a way to efficiently solve the hydrodynamical equations regardless of the wide range of physical conditions characterizing stellar interiors (e.g. density stratification and a wide range of Mach numbers).

The MUlti-dimensional Stellar Implicit Code (\MUSIC) follows a different approach by solving implicitly the fully compressible hydrodynamical equations \citep{viallet_towards_2011,2013_comparison}. The challenge for an implicit solver lies in the necessity of solving a large nonlinear system at each time step. In \cite{2013_comparison}, the best performance was obtained with Newton-Krylov methods, which combine the Newton-Raphson method with an iterative linear solver. It was shown that the iterative solver requires preconditioning in order to achieve fast convergence for large CFL$_\mathrm{hydro}$. In fact, within the framework of Newton-Krylov methods, the preconditioner is the crucial ingredient of the implicit solver. One of the important performance bottlenecks that was identified by the authors, particularly when considering three-dimensional calculations, is the inefficiency of ``black-box'' algebraic preconditioning techniques such as incomplete LU factorizations (ILU) for large CFL number computations. Furthermore, the memory requirement for the storage of the Jacobian matrix and the ILU factorization increases significantly in 3D, restricting the range of problems that can be addressed with such a method.

In this paper, we present an implicit solver which aims at overcoming these limitations. This is achieved by combining a Jacobian-free Newton-Krylov method with a preconditioner that is tailored for our physical equations, as described in more detail in Sect. \ref{Numerical}. In Sect. \ref{SI}, we design semi-implicit schemes that treat sound waves and thermal diffusion implicitly; in Sect. \ref{pbp}, we show how these semi-implicit schemes can be utilized to form an efficient preconditioner for the Newton-Krylov method. We present in Sect. \ref{results} results that illustrate the performance of the solver for idealized test problems and for stellar interiors. We conclude in Sect. \ref{conclusion}.





\section{Numerical Description}
\label{Numerical}


\MUSIC\ solves the equations describing the evolution of density, momentum, and internal energy, taking into account external gravity and thermal diffusion:

\begin{eqnarray}
\partial_t \rho &=& - \vec \nabla \cdot (\rho \vec u), \label{eq:cons1}\\
\partial_t ( \rho e ) &=& -\vec \nabla \cdot (\rho e \vec u) - p\vec \nabla \cdot \vec u + \vec \nabla \cdot (\chi \vec \nabla T), \label{eq:cons2} \\
\partial_t ( \rho \vec u ) &=& - \vec \nabla \cdot (\rho \vec u\otimes \vec u)-\vec \nabla p + \rho \vec g, \label{eq:cons3}
\end{eqnarray}

\noindent where $\rho$ is the density, $e$ the specific internal energy, $\vec u$ the velocity, $p$ the gas pressure, $T$ the temperature, $\vec g$ the gravitational acceleration, and $\chi$ the thermal conductivity. The system of equations is closed with an equation of state (EOS). For stellar interiors, these equations describe radiation-hydrodynamics in the diffusion limit. This is appropriate when the plasma is optically thick. In this case, the thermal conductivity due to photons is given by

\begin{equation}
\label{eq:chirad}
\chi = \frac{16 \sigma T^3}{3\kappa \rho},
\end{equation}

\noindent where $\kappa$ is the Rosseland mean opacity, and $\sigma$ the Stefan-Boltzmann constant. Furthermore, the EOS includes the contribution of radiation to the internal energy and pressure. Optionally, \MUSIC\ can solve the total energy equation in place of the internal energy equation:

\begin{equation}
\label{eq:Etot}
\partial_t(\rho \epsilon_t) = -\vec \nabla \cdot (\rho \epsilon_t \vec u + p \vec u) + \rho \vec u \cdot \vec g + \vec \nabla \cdot (\chi \vec \nabla T),
\end{equation}

\noindent where $\epsilon_t = e + \vec u^2/2$ is the specific total energy.

We follow the method of lines and perform the spatial discretization
independently of the time discretization \citep[see
  e.g.][]{leveque_finite_2007}. The spatial discretization is performed using a finite
volume method with staggered velocity components located at cell interfaces.
The numerical fluxes are calculated using an upwind, monotonicity
preserving method of \cite{vanLeer74}. The resulting scheme is second-order in space and total variation diminishing.

The ``conserved'' variables, for which the conservation equations (\ref{eq:cons1}$-$\ref{eq:cons3}) \& (\ref{eq:Etot}) are solved, are represented as the column vector $U=(\rho,\rho e\,\rho \vec u)$ when solving the internal energy equation, or $U=(\rho,\rho \epsilon_t,\rho \vec u)$ when solving the total energy equation. In \MUSIC, the unknowns are different from the conserved variables $U$, and are represented as the column vector $X = (\rho, e, \vec u)$ when solving the internal energy equation or $X = (\rho,\epsilon_t, \vec u)$ when solving the total energy equation.

The spatial discretization yields a system of ordinary differential equations:

\begin{equation}
\label{eq:mol}
\frac{dU}{dt} = R_U(X),
\end{equation}

\noindent where $R_U$ contains the flux differencing and source terms.

The time discretization is carried out using the second-order Crank-Nicolson method:

\begin{equation}
U(X^{n+1}) =  U(X^n) + \frac{\Delta t}{2}\big ( R_U(X^{n+1}) + R_U(X^n) \big ).
\end{equation}

\noindent We define the nonlinear residual function as

\begin{equation}
\label{eq:crank_nicholson}
F_U(X) = \frac{U(X) - U(X^n)}{\Delta t} - \frac{1}{2}\big ( R_U(X) + R_U(X^n) \big ),
\end{equation}

\noindent so that
\begin{equation}
\label{eq:newtimestep}
F_U(X^{n+1})=0
\end{equation}
\noindent defines the solution at time step $n+1$. Equation \eqref{eq:newtimestep} is solved with a Newton-Raphson method. At each Newton iteration, a linear problem of the form
\begin{equation}
\label{eq:fullsystem}
J\delta X = - F_U(X)
\end{equation}

\noindent must be solved, where $J = \partial F_U / \partial X$ is the Jacobian matrix.

The use of a Krylov iterative method like GMRES \citep{saad1986gmres} is a standard practice for solving Eq. \eqref{eq:fullsystem} when the matrix is large.
However, \cite{2013_comparison} find that, for CFL$_\mathrm{hydro} \gtrsim 10$, the iterative method requires preconditioning to remain effective. ILU factorizations can perform an adequate job at modest CFL numbers (CFL$_\mathrm{hydro} \lesssim 100$), but becomes inefficient at larger values of the CFL number. Furthermore, we find that such a preconditioning technique significantly increases the memory requirements. 

In this work, we adopt an approach in which the Jacobian matrix is never explicitly formed. Jacobian-free Newton-Krylov (JFNK) methods are a popular choice for the resolution of large nonlinear system of equations, see \cite{knoll_jacobian-free_2004} for a review. Since we do not form the Jacobian matrix, algebraic preconditioning techniques, such as ILU factorizations, have to be abandoned. Preconditioning the JFNK method, particularly when the CFL number is large, remains important for performance. One of the main goals of this paper is the design of an efficient preconditioner adapted specifically to the physics of stellar interiors.

From a physical point of view, at large CFL numbers (CFL$_\mathrm{hydro} \gtrsim 100$) waves propagate over a large portion, if not all of the computational domain during a single time step. Effectively, it is as if information propagates at an infinite speed, as in parabolic problems. This changes the mathematical nature of the problem, i.e. from hyperbolic to parabolic, resulting in numerical stiffness. To be efficient, the numerical method has to take this property into consideration. Multi-grid methods attempt to exploit this property by exchanging information between the large and small scales, see for example \cite{kifonidis_multigrid_2012}. We adopt another approach which consists of using legacy methods, known as semi-implicit (SI) schemes, as preconditioners for the Krylov solver. This strategy is known as ``physics-based preconditioning'' (PBP), as  the preconditioner is tailored for the physical problem, see e.g. \cite{mousseau_physics-based_2000,knoll_jacobian-free_2004,reisner_implicitly_2005,park_physics-based_2009}.  SI schemes treat implicitly only the terms that are responsible for numerical stiffness. 
The scheme derived this way is numerically stable for CFL$_\mathrm{hydro}$ greater than one, as the stiff physics is treated implicitly. However, the accuracy of the solution obtained by the SI scheme is usually quite poor due to the approximations involved in the derivation of the scheme (see Sect. \ref{SI}). Good accuracy can be achieved by embedding such a scheme within a Newton-Krylov method as a preconditioner. 
This work closely follows \cite{park_physics-based_2009}, adapting their method to our numerical scheme and physical equations.

\section{Semi-implicit schemes for gas dynamics}
\label{SI}

In this section, we derive SI schemes for the hydrodynamic equations (\ref{eq:cons1}$-$\ref{eq:cons3}) \& (\ref{eq:Etot})  which treat sound waves and thermal diffusion implicitly. The remaining terms (e.g. advection) are treated explicitly.  In this section, our only concern is to design schemes that are stable and inexpensive, rather than accurate. Later, we will use these schemes as preconditioners for a fully implicit and accurate method.

\subsection{Equations for $p$, $e$, and $\vec u$}

Our SI schemes are derived from the evolution equations for the primitive variables $V=(p, e, \vec u)$. These are

\begin{align}
& \partial_t p + \vec u \cdot \vec \nabla p =  - \Gamma_1 p \vec \nabla \cdot \vec u + (\Gamma_3-1) \vec \nabla \cdot \big ( \chi \vec \nabla T \big ), \label{eq:V_p} \\
& \partial_t e + \vec u \cdot \vec \nabla e  = - \frac{p}{\rho} \vec \nabla \cdot \vec u + \frac{1}{\rho} \vec \nabla \cdot \big ( \chi \vec \nabla T \big ), \label{eq:V_e} \\
& \partial_t \vec u + \vec u \cdot \vec \nabla \vec u = - \frac{1}{\rho}\vec \nabla p +  \vec g, \label{eq:V_u}
\end{align}

\noindent where $\Gamma_1$ and $\Gamma_3$ are the generalized adiabatic indices for a general equation of state. For a perfect gas without any internal degrees of freedom, these adiabatic indices reduce to $\Gamma_1 = \Gamma_3 = \gamma$, where $\gamma$ is the usual adiabatic index. The detailed derivation of the pressure equation, Eq. (\ref{eq:V_p}), is given in Appendix \ref{appendix:pressure_equation}.

To simplify the notation and without loss of generality, we will consider the one-dimensional version of these equations:

\begin{align}
& \partial_t p +  u \partial_x p =  - \Gamma_1 p \partial_x  u + (\Gamma_3-1) \partial_x \big ( \chi \partial_x T \big ), \label{eq:V_p_1D} \\
& \partial_t e + u \partial_x e  = - \frac{p}{\rho} \partial_x u + \frac{1}{\rho} \partial_x \big ( \chi \partial_x T \big ), \label{eq:V_e_1D} \\
& \partial_t u + u \partial_x u = - \frac{1}{\rho} \partial_x p - g, \label{eq:V_u_1D}
\end{align}

\noindent where we assumed that $\vec g = -g \vec e_x$, where $\vec e_x$ is a unity vector in the $x$-direction. Extension of the numerical scheme to higher dimensions is straightforward.

\subsection{Transformation matrices}
\label{transformation}

Having introduced the conserved variables $U$, the independent variables $X$, and the primitive variables $V$, we will need the transformation matrices $\partial V / \partial U$ and $\partial X / \partial V$, which are defined as:

\begin{align}
\delta V =& \frac{\partial V}{\partial U} \delta U, \\
\delta X =& \frac{\partial X}{\partial V} \delta V.
\end{align}

\noindent These matrices are given below for both the case of the internal and total energy equations, and the details of their derivation is postponed to Appendix \ref{appendix:transformation}.

\subsubsection{Internal energy equation}

When solving for the internal energy equation, $U = (\rho, \rho e, \rho u)$ and $X = (\rho, e, u)$, the transformation matrices take the following form:

\begin{equation}
\frac{\partial V}{\partial U}=
\begin{pmatrix}
\frac{\partial p}{\partial \rho} \big |_e - \frac{e}{\rho} \frac{\partial p}{\partial e} \Big |_\rho  & \frac{1}{\rho} \frac{\partial p}{\partial e} \Big |_\rho  & 0\\
-e/\rho & 1/\rho & 0\\
-u/\rho &  0 & 1/\rho\\
\end{pmatrix},
\end{equation}

\noindent and

\begin{equation}
\frac{\partial X}{\partial V} =
\begin{pmatrix}
\big ( \frac{\partial p}{\partial \rho} \big |_e \big )^{-1} & - \big ( \frac{\partial p}{\partial \rho} \big |_e \big )^{-1} \frac{\partial p}{\partial e} \Big |_\rho   & 0\\
0 & 1      & 0\\
0 & 0       &  1 \\
\end{pmatrix}.
\end{equation}

\noindent The required derivatives are those typically provided by EOS routines.

\subsubsection{Total energy equation}

When solving for the total energy equation, $U = (\rho, \rho \epsilon_t, \rho u)$ and $X = (\rho, \epsilon_t, u)$, the matrices take the form:

\begin{equation}
\frac{\partial V}{\partial U} =
\begin{pmatrix}
\frac{\partial p}{\partial \rho} \big |_e - \frac{\epsilon_t-u^2}{\rho} \frac{\partial p}{\partial e} \Big |_\rho  & \frac{1}{\rho} \frac{\partial p}{\partial e} \Big |_\rho  &  - \frac{u}{\rho} \frac{\partial p}{\partial e} \Big |_\rho\\
- (\epsilon_t-u^2)/\rho & 1/\rho & -u/\rho\\
-u/\rho &  0 & 1/\rho\\
\end{pmatrix},
\end{equation}

\noindent and

\begin{align}
\frac{\partial X}{\partial V} =&
\begin{pmatrix}
\big ( \frac{\partial p}{\partial \rho} \big |_e \big )^{-1} & - \big ( \frac{\partial p}{\partial \rho} \big |_e \big )^{-1} \frac{\partial p}{\partial e} \Big |_\rho   & 0\\
0 & 1 & u\\
0 & 0 & 1\\
\end{pmatrix}.
\end{align}

\subsection{SI scheme for sound waves}
\label{SI3}

We first design a SI scheme that treats sound waves implicitly. In Sect. \ref{appendix:sound_waves} we start with deriving the propagation equation for adiabatic acoustic fluctuations, which identifies the terms in the equations that need to be treated implicitly.

\subsubsection{Propagation equation for acoustic fluctuations}
\label{appendix:sound_waves}

In this section we neglect thermal diffusion and gravity. We linearize the 1D equations (\ref{eq:V_p_1D}) and (\ref{eq:V_u_1D}) around a uniform background state:

\begin{align}
\rho &= \rho_0 + \rho',\\
p &= p_0 + p',\\
 u &= 0 + u.
\end{align}

\noindent Keeping only linear terms in the perturbations, we obtain:

\begin{align}
\partial_t p'  =& - \Gamma_1 p_0 \partial_x u \label{eq:soundwaves_p}, \\
\partial_t u &  = - \frac{1}{\rho_0} \partial_x p' \label{eq:soundwaves_u}.
\end{align}

\noindent Next, we take the derivatives of Eq. (\ref{eq:soundwaves_p}) with respect to $t$ and Eq. (\ref{eq:soundwaves_u}) with respect to $x$. We substitute the result of the differentiation of Eq. (\ref{eq:soundwaves_u}) into the result of the differentiation of Eq. (\ref{eq:soundwaves_p}) to eliminate $\partial_{t x} u$ and obtain the wave equation that describes the adiabatic propagation of sound waves:

\begin{equation}
\partial_t^2 p' - a^2 \partial_x^2 p' = 0,
\end{equation}

\noindent where $a = \sqrt{\Gamma_1 p_0/\rho_0}$ is the adiabatic sound speed.

The terms on the r.h.s of Eqs. (\ref{eq:soundwaves_p}) \& (\ref{eq:soundwaves_u}) are responsible for the propagation of sound waves. To overcome the corresponding CFL limit, we will treat them implicitly in the following section.


\subsubsection{Pressure equation}

To treat sound waves implicitly, Sect. \ref{appendix:sound_waves} suggests that we treat the ``$-\Gamma_1 p \partial_ x u$'' term in the pressure equation (Eq. \ref{eq:V_p_1D}) implicitly, with a simple backward Euler method:

\begin{align}
\label{eq:pressure_intermediate}
 \frac{\delta p}{\Delta t} +  \Gamma_1 p^n \partial_x  u^{n+1}  = &- u \partial_x p\big |^n \notag \\
 & + (\Gamma_3-1) \partial_x \big ( \chi \partial_x T \big ) \big |^n.
\end{align}

\noindent Here $\delta p = p^{n+1} - p^{n}$, $n$ being the temporal index. We use Picard linearization in order to keep the scheme linear\footnote{Picard linearization refers to the fact that we write $p^n \partial_x  u^{n+1}$  instead of $p^{n+1} \partial_x u^{n+1}$ when applying the implicit discretization. This is an approximation, but it has the advantage of keeping the scheme linear in the new variables.}. All other terms in the equation are treated explicitly using the forward Euler method. Using Eq. (\ref{eq:soundwaves_u}), we approximate $u^{n+1}$ with

\begin{equation}
\label{eq:linearized_velocity}
u^{n+1} = u^{n} - \Delta t \frac{1}{\rho^n} \partial_x p^{n+1},
\end{equation}

\noindent which we substitute in Eq. (\ref{eq:pressure_intermediate}) to obtain

\begin{align}
\label{eq:pressure_implicit}
 \frac{\delta p}{\Delta t} - a^2 \Delta t \partial_x^2 p^{n+1}  = &- u \partial_x p\big |^n - \Gamma_1 p^n \partial_x  u^{n}  \notag \\
 & + (\Gamma_3-1) \partial_x \big ( \chi \partial_x T \big ) \big |^n,
\end{align}

\noindent where $a = \sqrt{\Gamma_1 p^n / \rho^n}$ is the adiabatic sound speed evaluated at time step $n$. The right hand side of Eq. (\ref{eq:pressure_implicit}) corresponds to the explicit discretization of the original equation, but on the left hand side a Laplacian operator illustrates the parabolic character of this equation.

\subsubsection{Internal energy equation}

We approach the internal energy equation, Eq. (\ref{eq:V_e_1D}), in the same way as the pressure equation.  Advection and thermal diffusion terms are discretized using an explicit scheme, and the compressional work is discretized using an implicit scheme. This produces:

\begin{equation}
\label{eq:energy_implicit0}
\frac{\delta e}{\Delta t} + \frac{p^n}{\rho^n} \partial_x u^{n+1}   = - u\partial_x e|^n +  \frac{1}{\rho^n}\partial_x \big ( \chi \partial_x T \big )|^{n},
\end{equation}

\noindent where $\delta e = e^{n+1} - e^{n}$. Again, we used Picard linearization to discretize the compressional work. We use again Eq. (\ref{eq:linearized_velocity}) to eliminate  $u^{n+1}$ in Eq. (\ref{eq:energy_implicit0}) to obtain

\begin{align}
\label{eq:energy_implicit}
\frac{\delta e}{\Delta t}  - \Delta t \frac{p^n}{(\rho^n)^2}  \partial_x^2 p^{n+1}  = & -  u\partial_x e|^n  - \frac{p^n}{\rho^n} \partial_x u^n \notag \\
&+  \frac{1}{\rho^n} \partial_x \big ( \chi \partial_x T \big )|^{n} .
\end{align}

\noindent The resulting equation is similar in form to the implicit version of the pressure equation, Eq. (\ref{eq:pressure_implicit}), as it contains the Laplacian of the pressure field.

\subsubsection{Velocity equation}

We discretize the pressure gradient in the velocity equation, Eq. (\ref{eq:V_u_1D}), implicitly, using Picard linearization. All remaining terms are discretized explicitly. We obtain

\begin{align}
\label{eq:u_implicit}
\frac{\delta u}{\Delta t} + \frac{1}{\rho^n} \partial_x p^{n+1} &= - u \partial_x u|^n - g,
\end{align}

\noindent where $\delta u = u^{n+1} - u^{n}$.

\subsubsection{``$\delta$-form'' of the equations}

By replacing $q^{n+1}=\delta q + q^n$ for all implicit terms in Eqs. (\ref{eq:pressure_implicit}, \ref{eq:energy_implicit}, \ref{eq:u_implicit}), rather than only those terms involving time derivatives, one obtains the following system of equations:

\begin{align}
\frac{\delta p}{\Delta t} - a^2 \Delta t \partial_x^2 \delta p &= - \tilde{F}_p(p^n), \label{eq:SI3_p}\\
\frac{\delta e}{\Delta t}  - \Delta t \frac{p^n}{(\rho^n)^2}  \partial_x^2 \delta p &= - \tilde{F}_e (e^n, p^n), \label{eq:SI3_e} \\
\frac{\delta u}{\Delta t} + \frac{1}{\rho^n} \partial_x \delta p &= - \tilde{F}_u (u^n, p^n),  \label{eq:SI3_u}
\end{align}

\noindent where we introduced the following residual functions:

\begin{align}
\tilde{F}_p (p) =& \frac{p - p^n}{\Delta t} - a^2 \Delta t \partial_x^2 p + u \partial_x p|^n + \Gamma_1 p^n \partial_x  u^{n}\notag \\
& -  \big ( \Gamma_3 -1 \big) \partial_x \big ( \chi \partial_x T \big )|^{n},\label{eq:res_p} \\
\tilde{F}_e(e, p) =&  \frac{e-e^n}{\Delta t} - \Delta t \frac{p^n}{(\rho^n)^2}  \partial_x^2 p +  u\partial_x e|^n  \notag \\
& + \frac{p^n}{\rho^n} \partial_x u^n -  \frac{1}{\rho^n} \partial_x \big ( \chi \partial_x T \big )|^{n}, \label{eq:res_e} \\
\tilde{F}_u(u, p) =& \frac{u - u^n}{\Delta t} + \frac{1}{\rho^n} \partial_x p + u \partial_x u|^n + g. \label{eq:res_u}
\end{align}

\noindent The solution at time $n+1$ satisfies $\tilde{F}_p(p^{n+1}) =0$, $\tilde{F}_e(e^{n+1}, p^{n+1})= 0$, and $\tilde{F}_u(u^{n+1}, p^{n+1}) = 0$.

\noindent We write the system in a matrix form:

\begin{equation}
\tilde{J}_V \delta V = - \tilde{F}_V (V^n),
\end{equation}

\noindent with $V = (p, e, u)$ and $\tilde{F}_V = (\tilde{F}_p, \tilde{F}_e, \tilde{F}_u)$. This formulation of the equations is known as the ``$\delta-$form''. The block structure of $\tilde{J}_V$ is:

\begin{equation}
\label{eq:SI3_blockstructure_V}
\tilde{J}_V =
   \begin{pmatrix} 
      \tilde{J}_{p,p} & 0 & 0 \\
       \tilde{J}_{e,p} & \tilde{J}_{e,e} & 0 \\
             \tilde{J}_{u,p} & 0 &\tilde{J}_{u,u}\\
   \end{pmatrix}.
\end{equation}

\noindent In this form, the system can be solved by operator splitting: the pressure equation (\ref{eq:SI3_p}) is first solved for $\delta p$, $\delta e$ is deduced from the internal energy equation (\ref{eq:SI3_e}), and $\delta u$ is deduced from the velocity equation (\ref{eq:SI3_u}). Note that in the energy equation, one can use the equality

\begin{equation}
\Delta t \partial_x^2 \delta p = \frac{1}{a^2} \Big ( \frac{\delta p}{\Delta t} + \tilde{F}_p \Big ),
\end{equation}

\noindent which is obtained from the pressure equation, rather than writing the Laplacian of $\delta p$ explicitly. In Sect. \ref{parabolic_system}, we discuss how we solve numerically the parabolic equation for $\delta p$.

\subsection{SI scheme for sound waves and thermal diffusion}
\label{sect:si_schemes}

When thermal diffusion is important, it can cause numerical stiffness. In this case, it also needs to be treated implicitly. This can be easily implemented in the framework of the previous section: all that is required is to treat thermal diffusion implicitly in the pressure and internal energy equations. Equation (\ref{eq:pressure_implicit}) now becomes:

\begin{align}
\label{eq:pressure_implicit2}
 \frac{\delta p}{\Delta t}  - a^2 \Delta t \partial_x^2 p^{n+1}  & - (\Gamma_3-1) \partial_x \big ( \chi^n \partial_x T^{n+1} \big )    \notag \\
 & = - u \partial_x p\big |^n - \Gamma_1 p^n \partial_x  u^{n},
\end{align}

\noindent and Eq. (\ref{eq:energy_implicit}) becomes:

\begin{align}
\label{eq:energy_implicit2}
\frac{\delta e}{\Delta t}  - \Delta t \frac{p^n}{(\rho^n)^2}  \partial_x^2 p^{n+1}  &-  \frac{1}{\rho^n} \partial_x \big ( \chi^n \partial_x T^{n+1} \big )  \notag \\
& = - u\partial_x e|^n  - \frac{p^n}{\rho^n} \partial_x u^n.
\end{align}

\noindent In both Eqs. (\ref{eq:pressure_implicit2}) and (\ref{eq:energy_implicit2}), we use Picard linearization to treat the diffusion term.

The new system for variables $V$ in $\delta$-form is

\begin{align}
\frac{\delta p}{\Delta t} &- a^2 \Delta t \partial_x^2 \delta p - (\Gamma_3-1) \partial_x \big ( \chi^n \partial_x \delta T \big ) = - \tilde{F}_p,\\
\frac{\delta e}{\Delta t}  &- \Delta t \frac{p^n}{(\rho^n)^2}  \partial_x^2 \delta p -  \frac{1}{\rho^n}\partial_x \big ( \chi^n \partial_x \delta T \big ) = - \tilde{F}_e, \\
\frac{\delta u}{\Delta t} &+ \frac{1}{\rho^n} \partial_x \delta p = - \tilde{F}_u,
\end{align}

\noindent where the residuals $\tilde{F}$ are unchanged, and given in Eqs. (\ref{eq:res_p}), (\ref{eq:res_e}) \& (\ref{eq:res_u}).

We use the linearized equation-of-state to express $\delta T$ as

\begin{equation}
\delta T = \frac{1}{c_v} \delta e + \frac{\partial T}{\partial \rho} \Big|_e  \delta \rho,
\end{equation}

\noindent where $c_v = \partial e / \partial T |_\rho$ is the specific heat capacity at constant volume\footnote{Here it is understood that partial derivatives are evaluated at time step $n$.}. In general, the contribution due to density fluctuations is much smaller\footnote{It is zero for a perfect gas, for which $e=e(T)$.} and we neglect them:

\begin{equation}
\delta T \approx \frac{1}{c_v} \delta e.
\end{equation}

\noindent This approximation is used to replace $\delta T$ with $\delta e$ in the previous system to obtain

\begin{align}
\frac{\delta p}{\Delta t} &- a^2 \Delta t \partial_x^2 \delta p -  (\Gamma_3-1) \partial_x \Big ( \frac{\chi^n}{c_v} \partial_x \delta e \Big ) = - \tilde{F}_p \label{eq:SI4_p_final}, \\
\frac{\delta e}{\Delta t}  &- \Delta t \frac{p^n}{(\rho^n)^2}  \partial_x^2 \delta p -  \frac{1}{\rho^n}\partial_x \Big ( \frac{\chi^n}{c_v} \partial_x \delta e \Big ) = - \tilde{F}_e \label{eq:SI4_e_final}, \\
\frac{\delta u}{\Delta t} &+ \frac{1}{\rho^n} \partial_x \delta p = - \tilde{F}_u \label{eq:SI4_u_final}.
\end{align}

\noindent We now have a system of two coupled parabolic equations, as seen from the block structure of the matrix $\tilde{J}_V$:

\begin{equation}
\label{eq:SI4_blockstructure_V}
\tilde{J}_V =
   \begin{pmatrix} 
      \tilde{J}_{p,p} & \tilde{J}_{p,e} & 0 \\
       \tilde{J}_{e,p} & \tilde{J}_{e,e} & 0 \\
      \tilde{J}_{u,p} & 0 & \tilde{J}_{u,u} \\
   \end{pmatrix}.
\end{equation}

\noindent The solution strategy of a system of equations in which thermal diffusion terms are treated implicitly is therefore more complicated than in the previous section: the two coupled parabolic equations (\ref{eq:SI4_p_final}) and (\ref{eq:SI4_e_final}) have to be solved jointly for $\delta e$ and $\delta p$, and finally $\delta u$ is obtained from Eq. (\ref{eq:SI4_u_final}).

\subsection{Numerical solution of the parabolic system}
\label{parabolic_system}

For both SI schemes presented previously, the numerical solution of a system of linear parabolic equations is required. This is accomplished in \MUSIC\ using the Trilinos library \citep[see][]{heroux_overview_2005}. Specifically, \MUSIC\ uses the iterative linear solver GMRES implemented in the package {\sf AztecOO} to solve the parabolic system. The convergence of the linear solver is checked based on the criterion:

\begin{equation}
\label{eq:forcing_term2}
|| P x - b||_2 < \eta' ||b||_2,
\end{equation}

\noindent where $P$ is the system matrix, $x$ the solution vector, $b$ the
r.h.s. of the linear system, and $\eta'$ controls the accuracy of the solution. When setting $\eta' \leq 10^{-6}$, we find that the preconditioner has the same performances as when we use a direct solver to solve the parabolic system. However, in practice it is not necessary to solve the parabolic problem with such accuracy, as the preconditioner is only meant to provide an approximate solution of the problem. The results presented in this work were obtained by adopting a value $\eta'=10^{-4}$. This value ensures an accuracy that is sufficient for the purpose of preconditioning. It is possible that the performance could be improved by adopting even larger values of $\eta'$, as the decrease in the quality of the preconditioner could be mitigated by the decrease in its computational cost. This is left for future investigation. A multi-level preconditioner is applied to speed up convergence of the linear solver. We use the {\sf ML} package of the {\sf Trilinos} library to setup a multi-level preconditioner \citep{ml-guide}. Our preconditioner is based on the default parameters provided for the smoothed-aggregation setup in the {\sf ML} package (parameters set ``SA'') with the following two modifications. First, instead of using the default method to estimate the eigenvalues of the matrix we use the 1-norm of the matrix. The default method of estimating the eigenvalues used a method based on a conjugate-gradient solution of the system, seeded by a random vector. This random vector caused simulations continued after restarting to differ from simulations without the restart, removing the ability to reproduce results. Second, we reduce the damping factor for the precondition from the default of 1.33 to 1.2, which in our case, results in fewer iterations for the parabolic solver to converge.




\subsection{Time-stepping with SI schemes}
\label{si_timestepping}

The SI schemes designed in this section can be used as time-stepping methods to solve Eqs. (\ref{eq:cons1}$-$\ref{eq:cons3}) \& (\ref{eq:Etot}). The time-marching algorithm is:

\begin{enumerate}
\item Given the solution at time step $n$, $X^n$, compute $F_U(X^n)=-R_U(X^n)$ (see Eq. \ref{eq:crank_nicholson});
\item Transform $F_U$ into a residual for the primitive variables $V$:
\begin{equation}
\tilde{F}_V = \frac{\partial V}{\partial U} F_U;
\end{equation}
\item Compute $\tilde{J}_V(V^n)$ corresponding to the desired SI scheme and solve
\begin{equation}
\tilde{J}_V \delta V = - \tilde{F}_V
\end{equation}
for $\delta V$;
\item Transform $\delta V$ into $\delta X$:
\begin{equation}
\delta X = \frac{\partial X}{\partial V} \delta V;
\end{equation}
\item Set $X^{n+1} = X^n + \delta X$.
\end{enumerate}

\noindent The scheme is linear (we used Picard linearization to deal with nonlinear terms) and only first-order in time (we used the forward/backward Euler methods). However the CFL limit is less restrictive as the terms associated with acoustic fluctuations were discretized implicitly. Similarly, thermal diffusion does not imply any stability restriction on the time step if the second SI scheme is used. However, since the advective terms were discretized explicitly, a time step restriction based on the flow speed remains.

\subsection{2D isentropic vortex test}
\label{si:tests}


\begin{figure*}[t] 
   \centering
   \parbox{0.45\linewidth}{\centering \includegraphics[width=0.8\linewidth, trim= 60 0 60 20]{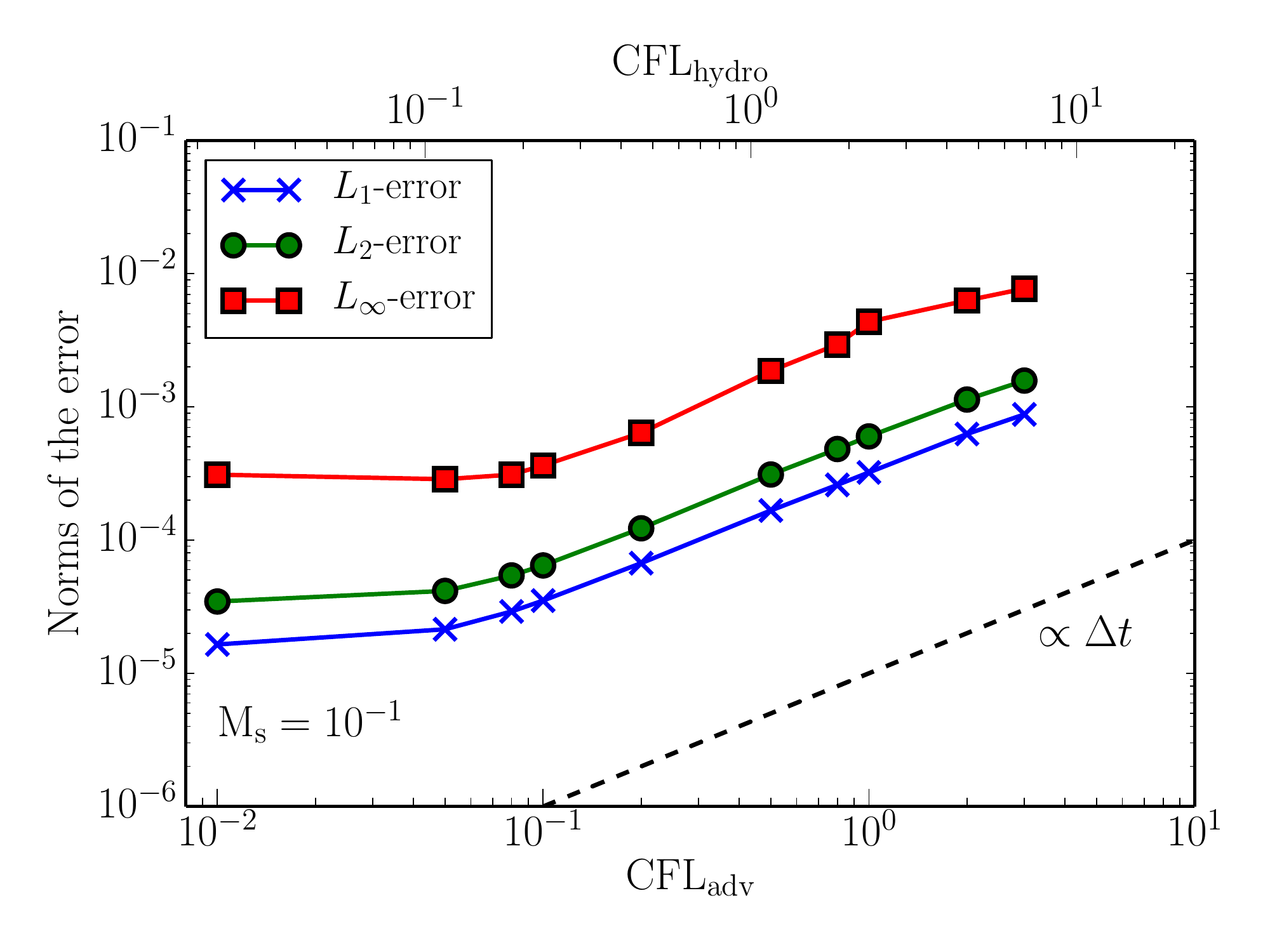}}
   \parbox{0.45\linewidth}{\centering \includegraphics[width=0.8\linewidth, trim= 60 0 60 20]{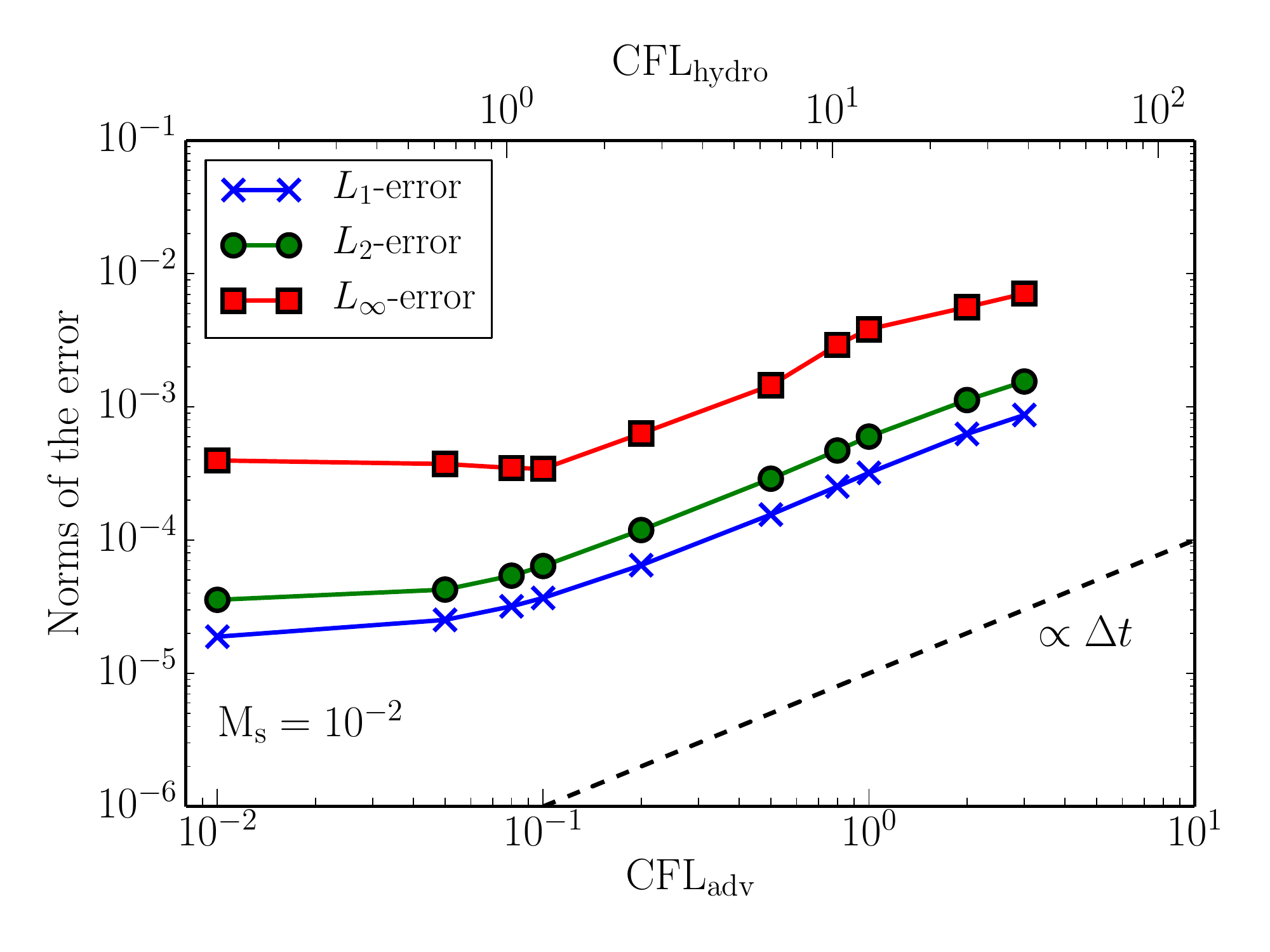}}
   \parbox{0.45\linewidth}{\centering \includegraphics[width=0.8\linewidth, trim= 60 0 60 20]{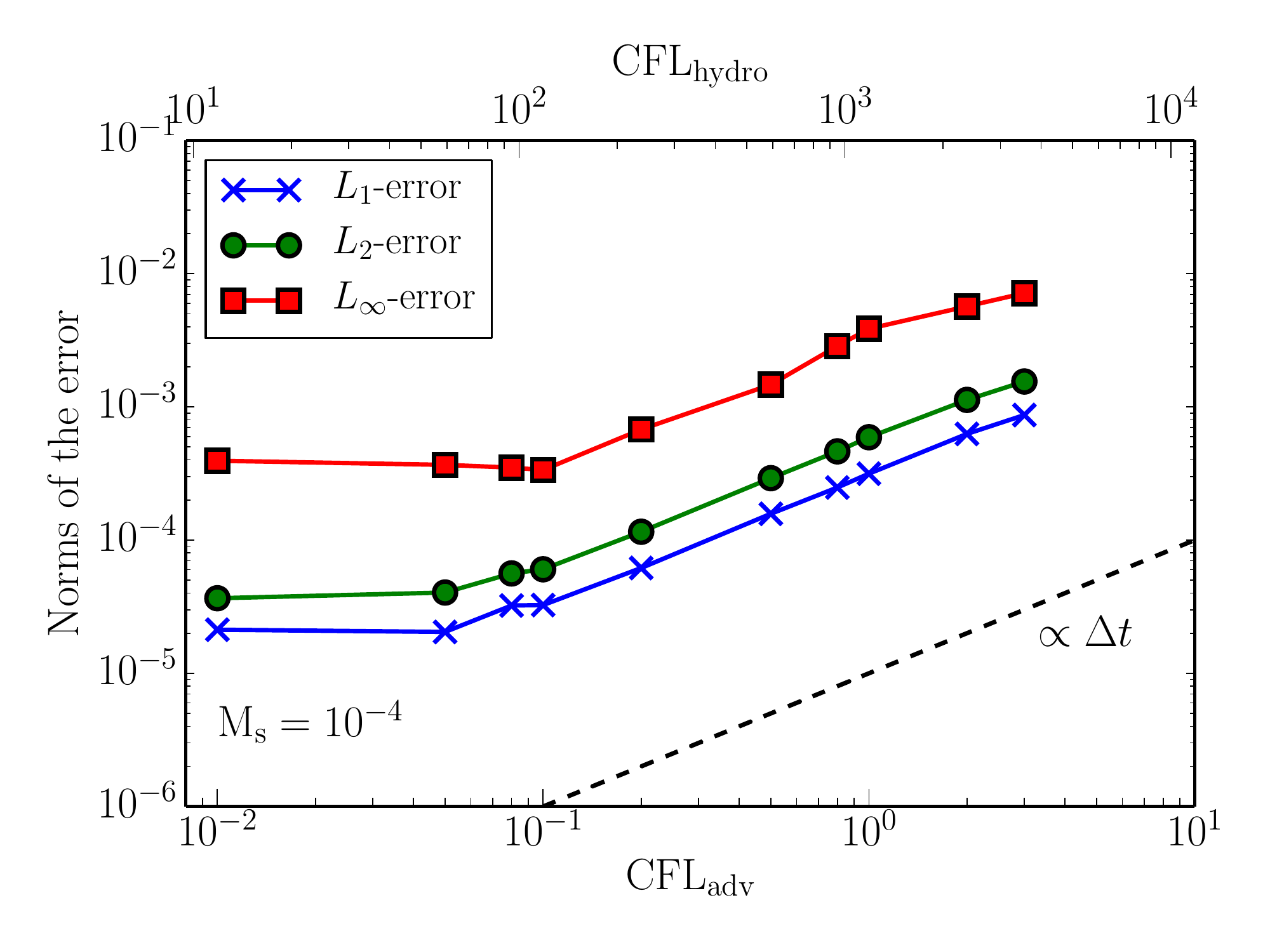}}
   \parbox{0.45\linewidth}{\centering \includegraphics[width=0.8\linewidth, trim= 60 0 60 20]{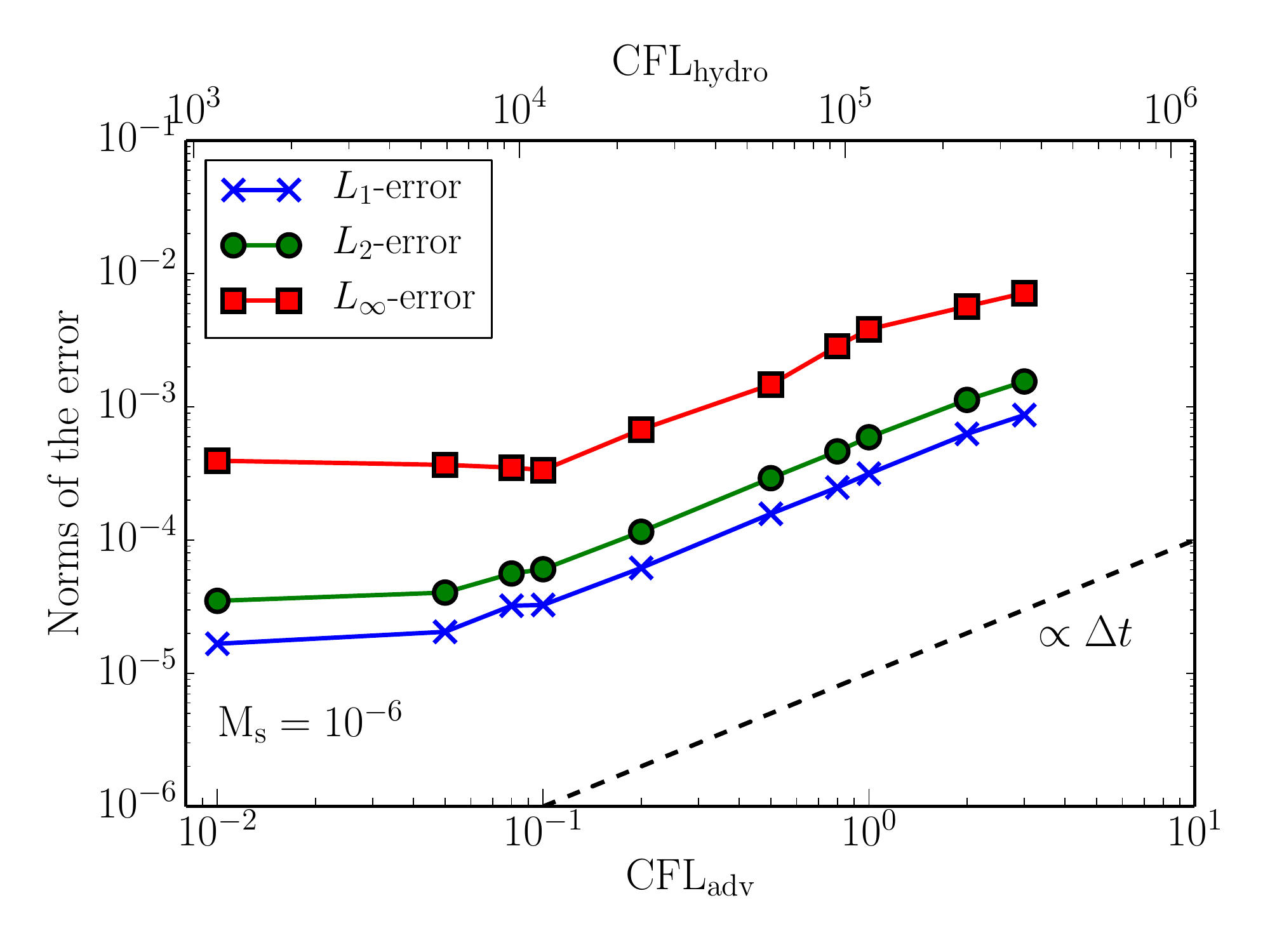}}
   \caption{Convergence tests of the SI scheme treating sound waves implicitly - advection of an isentropic vortex at different Mach numbers. The continuous lines show the norm of the errors measured in the velocity component parallel to the direction of advection.}
   \label{fig:test_si_vortex_adv}
\end{figure*}

In this section, we test the SI scheme for sound waves. We use the isentropic vortex advection test originally proposed in \cite{yee_entropy_2000}, and we adopt a setup similar to the one used by \cite{kifonidis_multigrid_2012} and \cite{2013_comparison} to test the accuracy of the SI scheme.

The initial state consists of an isentropic vortex (i.e. zero entropy perturbation) embedded in an uniform flow of norm $u_\infty=1$. We use a Cartesian system of coordinates where the $x$-axis is taken in the direction of the flow. The vortex corresponds to the following perturbations in the state variables:

\begin{align}
(\delta u, \delta v) &= \frac{\beta}{2\pi} \mathrm{e}^{\frac{1-r^2}{2}} (-y, x),\\
\delta T &= - \frac{(\gamma-1) \beta^2}{8 \gamma \pi^2} \mathrm{e}^{1-r^2}, \label{eq:dt}
\end{align}

\noindent where $r=\sqrt{x^2+y^2}$, $T=p/\rho$ (we set the gas constant $R=1$ and work with dimensionless quantities), $\gamma$ is the adiabatic index, and $\beta$ the vortex strength. We use $\gamma=1.4$ and $\beta=0.75$, with initial conditions

\begin{align}
\rho &= (T_\infty + \delta T)^{\frac{1}{\gamma-1}}, \\
u &= u_\infty + \delta u, \\
v &= \delta v, \\
e &= \frac{\rho^{\gamma-1}}{\gamma-1},
\end{align}

\noindent where the subscript $\infty$ indicates the background value. The sound speed of the background is $c_\infty = \sqrt{\gamma T_\infty}$. The maximum velocity of the vortex is $v_\mathrm{max} = \max || \delta \vec u|| = \beta/2\pi$. We define the vortex Mach number as $M_s=v_\mathrm{max}/c_\infty = \beta/( 2\pi \sqrt{\gamma T_\infty})$. By varying $T_\infty$, we change the Mach number of the flow. We consider $T_\infty = 1, 10^2, 10^6, 10^{10}$, which corresponds to $M_s = 10^{-1}$, $10^{-2}$, $10^{-4}$, $10^{-6}$ respectively.

The computations are performed on a 2D Cartesian domain $[-4,4]\times[-4,4]$. Initially, the vortex is centered on the origin. The vortex is advected until $t=0.4$. The exact solution corresponds to the initial vortex profile being shifted by a distance equal to $0.4$ in the $x$ direction. To test the accuracy of the scheme, we compare the velocity in the direction of advection, $u$, with the expected analytic solution $u^0$. \cite{kifonidis_multigrid_2012} and \cite{2013_comparison} used the density field to monitor the error, but here the background density is changed when $T_\infty$ is changed, which is not the case for the velocity field. We monitor three different norms of the error:

\begin{align}
& L_1\mathrm{-error:\ } || u - u^0 ||_1 = \frac{1}{N_x N_y} \sum_{i,j} | u_{i,j} - u^0_{i,j} |, \\
& L_2\mathrm{-error:\ } || u - u^0 ||_2 = \sqrt{ \frac{1}{N_x N_y} \sum_{i,j} ( u_{i,j} -u^0_{i,j} )^2 }, \\
& L_\infty \mathrm{-error:\ } || u - u^0 ||_\infty = \max_{i,j} | u_{i,j} - u^0_{i,j} |,
\end{align}

\noindent where $N_x$, $N_y$ are the grid dimensions, and the indices $i$, $j$ range over the simulation grid.

We introduce the advective CFL number:

\begin{equation}
\mathrm{CFL}_\mathrm{adv} = \frac{u_\infty \Delta t}{\Delta x},
\end{equation}

\noindent where $\Delta t$ is the time step and $\Delta x$ the mesh spacing. For a vortex advected at $u_\infty$, the advective CFL number provides a measure of the number of grid cells crossed per time step $\Delta t$.

To characterise the accuracy of the SI scheme, we perform a temporal convergence study. The resolution of the domain is set to $64^2$ and we choose different time steps in order to cover a broad range of CFL$_\mathrm{adv}$, between  $\sim 10^{-2}$ and 3. This corresponds to CFL$_\mathrm{hydro}$ as large as $4\times 10^5$. Later, we will compare the result with a more accurate time integration method. We do not study convergence with spatial resolution here, as our spatial method remains the same in all schemes presented in this work, and is unchanged as compared to previous publications. A spatial resolution study is presented in \cite{viallet_towards_2011}.

We evolve the isentropic vortex varying both the Mach number and CFL$_\mathrm{adv}$, and monitor the numerical errors. We expect two behavioral regimes. At low values of CFL$_\mathrm{adv}$, the error should be approximately independent of the time step, as the spatial error dominates. At higher values of CFL$_\mathrm{adv}$, the temporal error should dominate, and be proportional to $\Delta t$ as the SI scheme is first-order in time. The results are presented in Fig. \ref{fig:test_si_vortex_adv}. The expected behavior is recovered, although the flat regime at low values of the time step is not clearly seen. Temporal truncation errors remain significant for small time steps, as a result of the approximations introduced when designing the SI scheme. The first-order character of the temporal discretization appears clearly for larger values of the time step. However, the most important conclusion is that the numerical error is independent of the Mach number. Effectively, we achieved our goal of designing a scheme that is independent of the stiffness of the background pressure field. We stress that this behavior is observed over a range of Mach numbers spanning five orders of magnitude.

Finally, although we successfully removed the stability constraint on the time step caused by sound waves, there is still a CFL-like condition based on the advective velocity. Such a stability limit is not evident from Fig. \ref{fig:test_si_vortex_adv}, as only a few models are computed for the largest time steps. Empirically, we determined that the SI scheme becomes unstable for CFL$_\mathrm{adv} \gtrsim 0.2$.



\section{Jacobian-free Newton-Krylov method and physics-based preconditioning}
\label{pbp}

\subsection{Newton-Krylov method}
\label{pbp:nk}

To solve the nonlinear system of equations, $F_U(X^{n+1})=0$, resulting from our fully implicit method we perform Newton-Raphson iterations. The Newton-Raphson procedure is initiated by taking an initial guess for the solution, typically $X^{(0)} = X^n$. At the $k$-th Newton-Raphson iteration, the solution of a linear system is required:

\begin{equation}
\label{eq:jacobian_system}
J^{(k)} \delta X^{(k)} = - F_U(X^{(k)}),
\end{equation}

\noindent where $\delta X^{(k)} = X^{(k+1)} - X^{(k)}$. The variable $X^{(k)}$ is the solution at iteration $k$, and

\begin{equation}
J^{(k)} = \frac{\partial F_U}{\partial X}(X^{(k)})
\end{equation}

\noindent is the Jacobian matrix at iteration $k$.

\noindent The components of $\delta X$ and $F_U$ can have considerably different numerical values as they represent different physical quantities in different units. For instance, densities can have typical values around $10^{-4}$ g/cc, whereas specific internal energies have values around $10^{14}$ erg/g. Also, due to the stratification of stellar interiors, some variables, such as the density, can vary by several orders of magnitude throughout the domain. Such a wide range of values can cause numerical difficulties due to round-off errors. Therefore, before the system (\ref{eq:jacobian_system}) can be solved, it is necessary to scale it. We introduce two diagonal matrices $L$ and $R$ to scale Eq. (\ref{eq:jacobian_system}):

\begin{equation}
\label{eq:jacobian_system_norm}
\left( L^{-1} J^{(k)} R \right ) \left (R^{-1} \delta X^{(k)} \right)= - L^{-1} F_U(X^{(k)}).
\end{equation}

\noindent As $L$ and $R$ are diagonal matrices, we use the same symbol to represent their diagonal entries as a vector. The size of these vectors is equal to the number of variables multiplied by the number of cells. Each cell is treated in the same way, and the definitions of $R$ and $L$ only differ for different variables:

\begin{equation}
\begin{split}
&L_\rho = \rho^{(k)}, \\
&L_e = \rho^{(k)} e^{(k)}, \\
&L_u = \rho^{(k)} \max(|u^{(k)}|, \alpha_1 c_s^{(k)}),
\end{split}
\hspace{0.5cm}
\begin{split}
&R_\rho = \rho^{(k)}, \\
&R_e = e^{(k)}, \\
&R_u = \max(|u^{(k)}|, \alpha_2 c_s^{(k)}),
\end{split}
\nonumber
\end{equation}

\noindent where $c_s^{(k)}$ is the adiabatic sound speed computed from the solution at iteration $k$. $R$ represents the typical value of the unknown vector $X^{(k)}$, and attempts to remove both the effects of units and stratification. We follow a similar idea for $L$ and use the typical value of the conserved variables to scale the residual vector $F_U$. However, as velocities can be arbitrarily small, it is necessary to introduce a minimum velocity, here measured relative to the sound speed using the parameters $\alpha_1,\alpha_2$ in the definitions of $L$ and $R$. The work described in \cite{viallet_towards_2011} and \cite{2013_comparison} used $\alpha_1 = \alpha_2 = 1$. After testing, we found that $\alpha_1 = 10^{-5}$ and $\alpha_2=1$ gives good performance for a wide range of Mach numbers, typically $10^{-6} \lesssim M_s \lesssim 10^{-1}$, see discussion in Sect. \ref{results}.


The Newton-Raphson procedure is terminated when the relative corrections fall below a certain value $\epsilon$:

\begin{equation}
\label{eq:nr_terminate}
|| R^{-1} \delta X^{(k)}||_\infty < \epsilon.
\end{equation}

\noindent In \cite{2013_comparison}, it was shown that the nonlinear tolerance $\epsilon$ has to be chosen small enough so that the truncation errors of the scheme dominate the numerical error. We follow their recommendation and set $\epsilon=10^{-6}$. Finally, if Eq. (\ref{eq:nr_terminate}) is already fulfilled at the first iteration, we enforce a second Newton-Raphson iteration. For the sake of clarity, we drop from now on the superscript $k$ of the outer nonlinear iteration of the Newton-Raphson procedure, and we do not carry the scaling matrices $L$ and $R$ in the notation in the rest of the paper.

We use the GMRES method to solve iteratively Eq. (\ref{eq:jacobian_system_norm}). We start from an initial guess $\delta X_{0}$, and we define the initial residual as $r_0 = -F_U(X) - J \delta X_{0}$. In practice, we choose $\delta X_{0} = 0$ so that $r_0 = -F_U(X)$.
At the $p$-th iteration, the GMRES method seeks an approximation $\delta X_p$ of the solution by solving a minimization problem in the $p$-th Krylov space $\mathcal{K}_p$ of $J$:

\begin{equation}
\delta X_p \in \mathcal{K}_{p}(J) = \mathrm{span} \Big ( r_0, J r_0, \dots, J^{p-1} r_0 \Big).
\end{equation}

\noindent The dimension of the search space increases at each iteration until convergence is achieved.
 The convergence of the linear solver is tested with the criterion

\begin{equation}
\label{eq:forcing_term}
|| J \delta X_p + F_U(X)||_2 < \eta ||F_U(X)||_2,
\end{equation}

\noindent where $\eta$ is a parameter that determines the accuracy of the solution. Typical values of $\eta$ that are used in this paper are $\eta = 10^{-2}$ and $\eta = 10^{-4}$, see discussion in Sect. \ref{results}.

\subsection{Jacobian-free approach}
\label{pbp:jf}

To build successive Krylov spaces, the GMRES algorithm computes the action of the Jacobian matrix on a vector. This is the only use of the Jacobian operator, and we take advantage of the fact that this operation can be approximated by finite-differencing:

\begin{equation}
\label{eq:jacobian_free}
J(\vec u) \vec v \approx \frac{F(\vec u + \delta \vec v) - F(\vec u)}{\delta},
\end{equation}

\noindent where $\delta$ is a small number. We rely on the implementation of matrix-free operators available from the {\sf Trilinos} package {\sf NOX}. This package contains two preset options for calculating $\delta$:

\begin{equation}
  \label{pert_old}
  \delta = \lambda \left ( \lambda + \frac{||\vec u||}{||\vec v||}\right ),
\end{equation}
and,
\begin{equation}
  \label{pert_new}
  \delta = \lambda \left(\frac{10^{-12}}{\lambda} + \frac{\left|\vec u\cdot \vec v\right|}{\vec v \cdot \vec v}\right)\textrm{sign}\left(\vec{u} \cdot \vec{v}\right).
\end{equation}

\noindent In both cases, $\lambda$ is a small parameter with a default value of $10^{-6}$. The standard choice in \MUSIC\ is to use Eq. (\ref{pert_old}), as it gives the best results (see discussion in Sect. \ref{results}).

In this Jacobian-free approach, the Jacobian matrix is not needed explicitly, lowering the memory cost of the scheme. Instead, computing the action of the Jacobian on a given vector requires one evaluation of the nonlinear residual in Eq. (\ref{eq:jacobian_free}), assuming that $F(\vec u)$ has been already computed and stored.

When $J$ has a large condition number, the Krylov method fails to converge in an acceptable number of iterations (a few dozen) as the Krylov space is dominated by the direction of the eigenvector associated with the largest eigenvalue. In such cases, preconditioning is necessary. In this work, we use the SI schemes presented in Sect. \ref{SI} as preconditioners for the Krylov method. This is detailed in the next section.

\subsection{Right-preconditioning of GMRES with SI schemes}


Right-preconditioning of system (\ref{eq:jacobian_system_norm}) corresponds to solving the equivalent system:

\begin{align}
\big( J M^{-1} \big)  \delta X' &= - F_U(X), \label{eq:lin_sys_rp} \\
M \delta X &= \delta X' , \label{eq:lin_sys_rp2}
\end{align}

\noindent where $M$ is the preconditioning matrix. $\delta X'$ is an intermediate solution vector, which once known, is used to find the solution $\delta X$. If the preconditioning matrix is a good approximation of $J$, i.e. $J M^{-1}$ has a low condition number, the Krylov space of $J M^{-1}$ is better suited to construct an approximation of the solution

\begin{equation}
\delta X'_p \in \mathcal{K}_{p}(J M^{-1} ) = \mathrm{span} \Big ( r_0, J M^{-1}  r_0, \dots, \big ( JM^{-1}\big)^{p-1} r_0 \Big).
\end{equation}

\noindent Once a suitable solution $\delta X'$ has been found in the search space, based on the same convergence criterion as \eqref{eq:forcing_term}, a final linear system, Eq. \eqref{eq:lin_sys_rp2}, has to be solved to get the actual solution $\delta X$.

The key part of the right-preconditioning process is the application of $J M^{-1}$ on a Krylov vector $v$, provided by GMRES. This operation is required at each iteration to build the successive Krylov spaces. In right-preconditioning, $J M^{-1} v$ is computed in two steps:

\begin{enumerate}
\item Solve $M w = v$ for $w$;
\item Apply $J$ to $w$.
\end{enumerate}

\noindent The first step requires the inversion of a linear system; the second step requires the action of the Jacobian on the vector $w$ and is approximated by a finite-difference formula (Jacobian-free approach).

The basic idea of physics-based preconditioning is to interpret the system $M w = v$ in step 1 above as a system corresponding to a linear time-stepping scheme written in $\delta$-form:

\begin{equation}
\label{eq:SI_prec}
M w = v \Leftrightarrow M \delta X = - G(X),
\end{equation}

\noindent where ($M$,$G$) describes a numerical scheme that approximates the full nonlinear scheme ($J$,$F$). Another way to understand physics-based preconditioning is that Eq. (\ref{eq:SI_prec}) defines a mapping $M$ from residuals to perturbations $\delta X$. Therefore, the Jacobian matrix is always applied to a $\delta X$ to yield a residual vector $F_U$ which is used to build Krylov spaces. 

The SI schemes designed in Sect. \ref{SI} are good candidates for the scheme in Eq. (\ref{eq:SI_prec}). These schemes provide a good approximation of the solution (i.e. $M \sim J$), and most importantly they remove the numerical stiffness by solving the stiff physics (sounds waves and thermal diffusion) implicitly. Physics-based preconditioner therefore ``injects'' physical insight at the heart of the linear method, improving its convergence. However, the Krylov vector is a residual for variables $U$, and it needs to be transformed into a residual for variables $V$ before a SI scheme can be used. Furthermore, the SI scheme provides $\delta V$, which needs to be transformed into $\delta X$ before $J$ can be applied. As in the time-stepping algorithm described in \ref{si_timestepping}, we use the matrices derived in Sect. \ref{transformation} to do these transformations. The complete algorithm to use the SI as a preconditioner is:

\begin{enumerate}
\item[--] Input: the GMRES method provides a vector $v \in \mathcal{K}_{p}$. $v$ can be interpreted as a residual vector for the conservative variables $U$, which we denote $F_U$;
\item Transform $F_U$ into a residual for the primitive variables $V$:
\begin{equation}
\tilde{F}_V = \frac{\partial V}{\partial U} F_U;
\end{equation}
\item Apply the SI scheme to get $\delta V$:
\begin{equation}
\tilde{J}_V \delta V = - \tilde{F}_V;
\end{equation}
\item Transform $\delta V$ into $\delta X$:
\begin{equation}
\delta X = \frac{\partial X}{\partial V} \delta V;
\end{equation}
\item Compute $J\delta X$ using the Jacobian-free method.
\item[--] Output: vector $J \delta X$, which is provided to GMRES to build successive Krylov spaces.
\end{enumerate}


We note that the scheme is not fully matrix free: the SI scheme requires the resolution of a linear problem for which the matrix is explicitly formed and stored. However, thanks to the simplifications made in deriving the SI scheme, the matrix system is significantly smaller and more sparse than the Jacobian matrix. This keeps memory demand low.

In the remainder of the paper, the combination of the Jacobian-free Newton-Krylov method with physics-based preconditioning presented in this section will be referred to as the JFNK+PBP method.



\begin{figure*}[t] 
   \centering
   \parbox{0.45\linewidth}{\centering \includegraphics[width=0.8\linewidth, trim= 60 0 60 20]{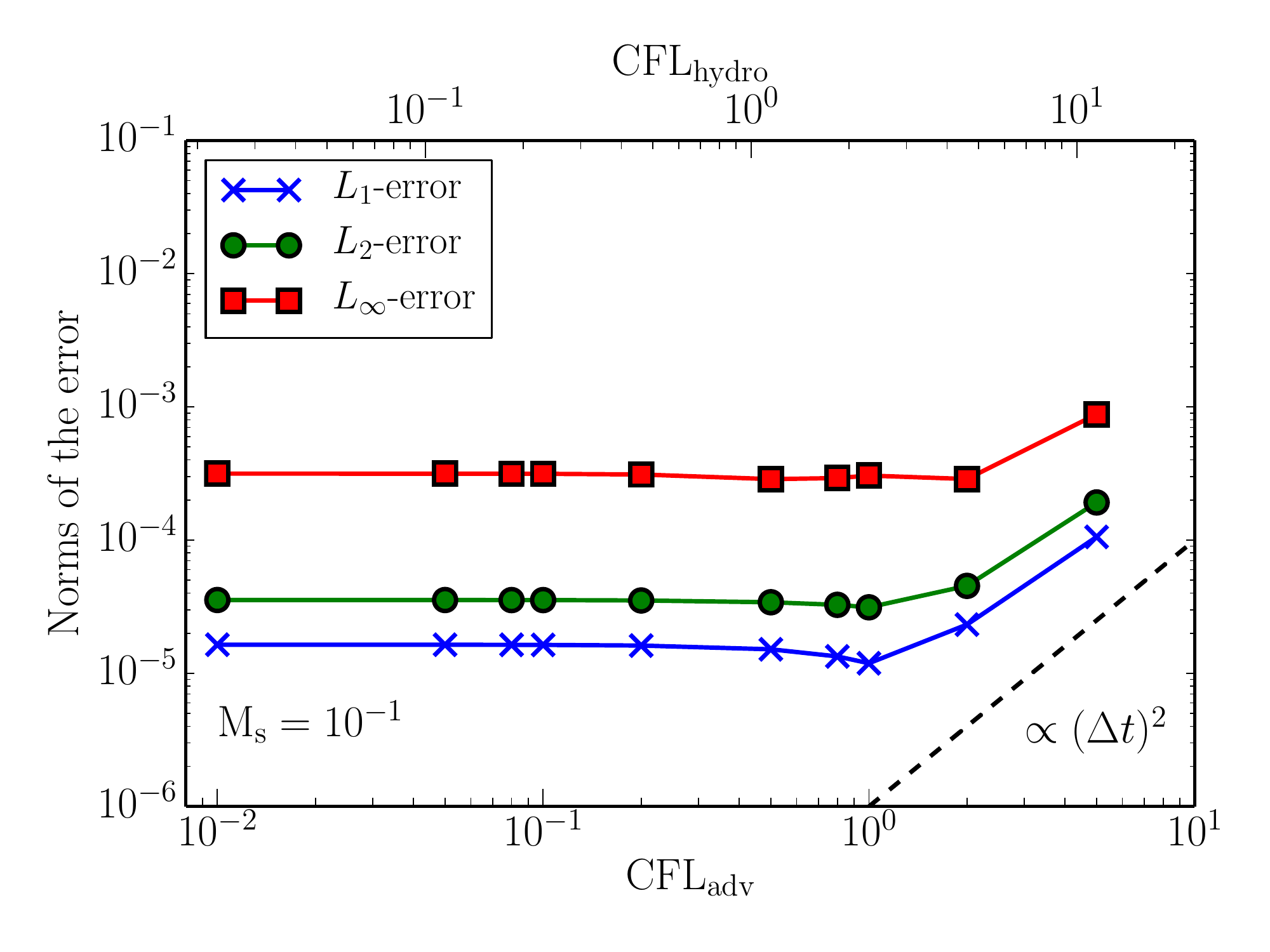}}
   \parbox{0.45\linewidth}{\centering \includegraphics[width=0.8\linewidth, trim= 60 0 60 20]{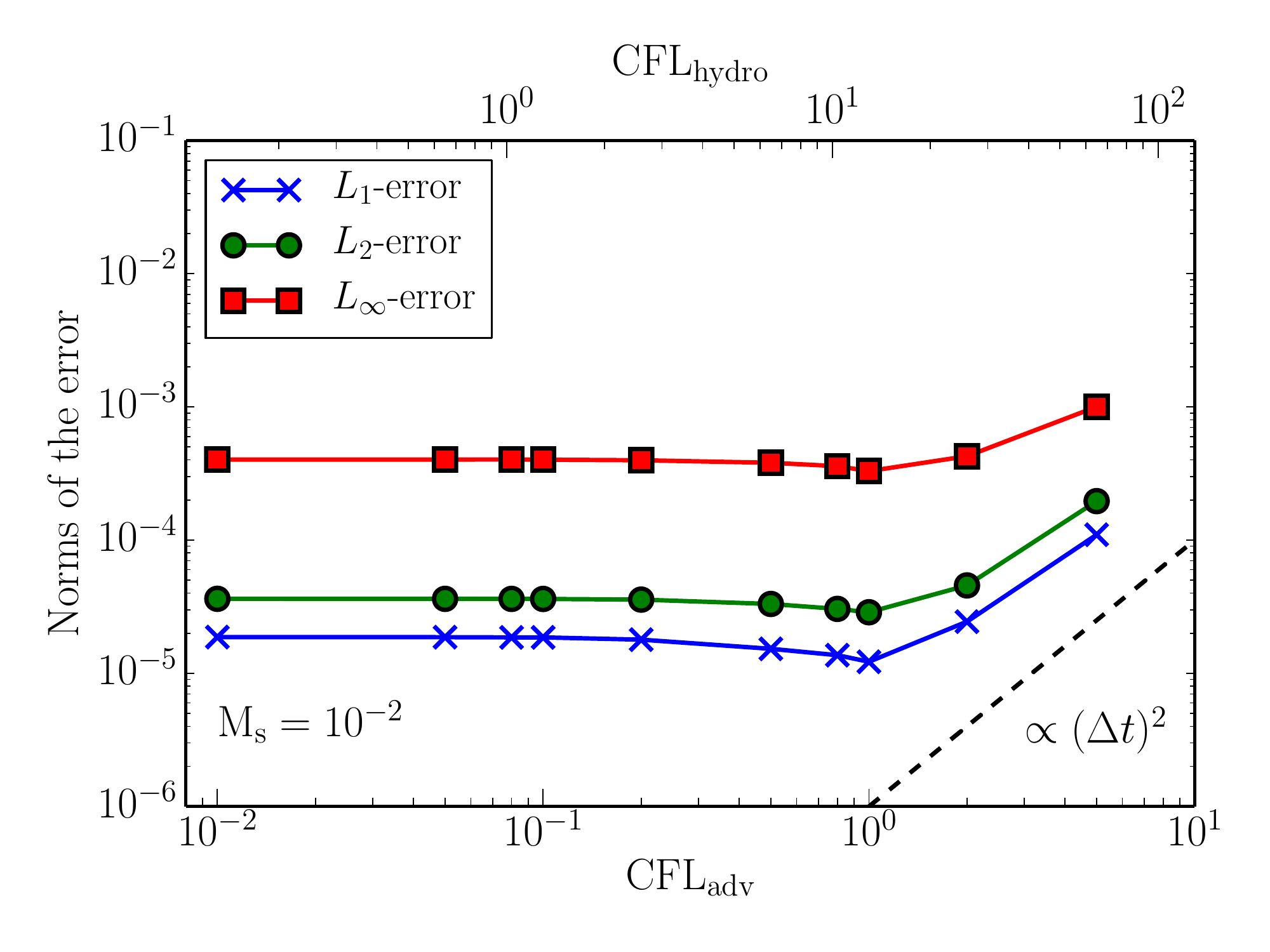}}
   \parbox{0.45\linewidth}{\centering \includegraphics[width=0.8\linewidth, trim= 60 0 60 20]{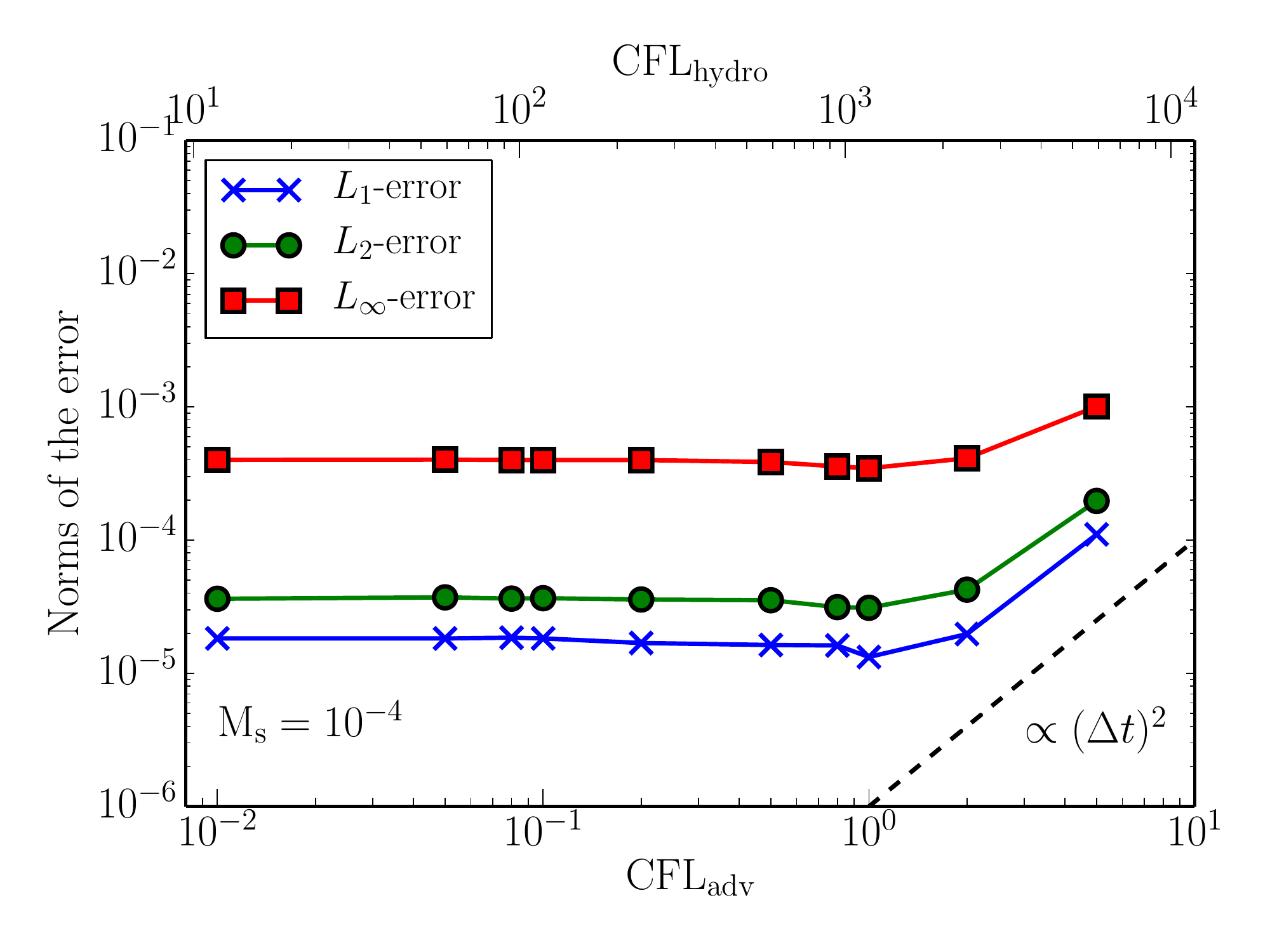}}
   \parbox{0.45\linewidth}{\centering \includegraphics[width=0.8\linewidth, trim= 60 0 60 20]{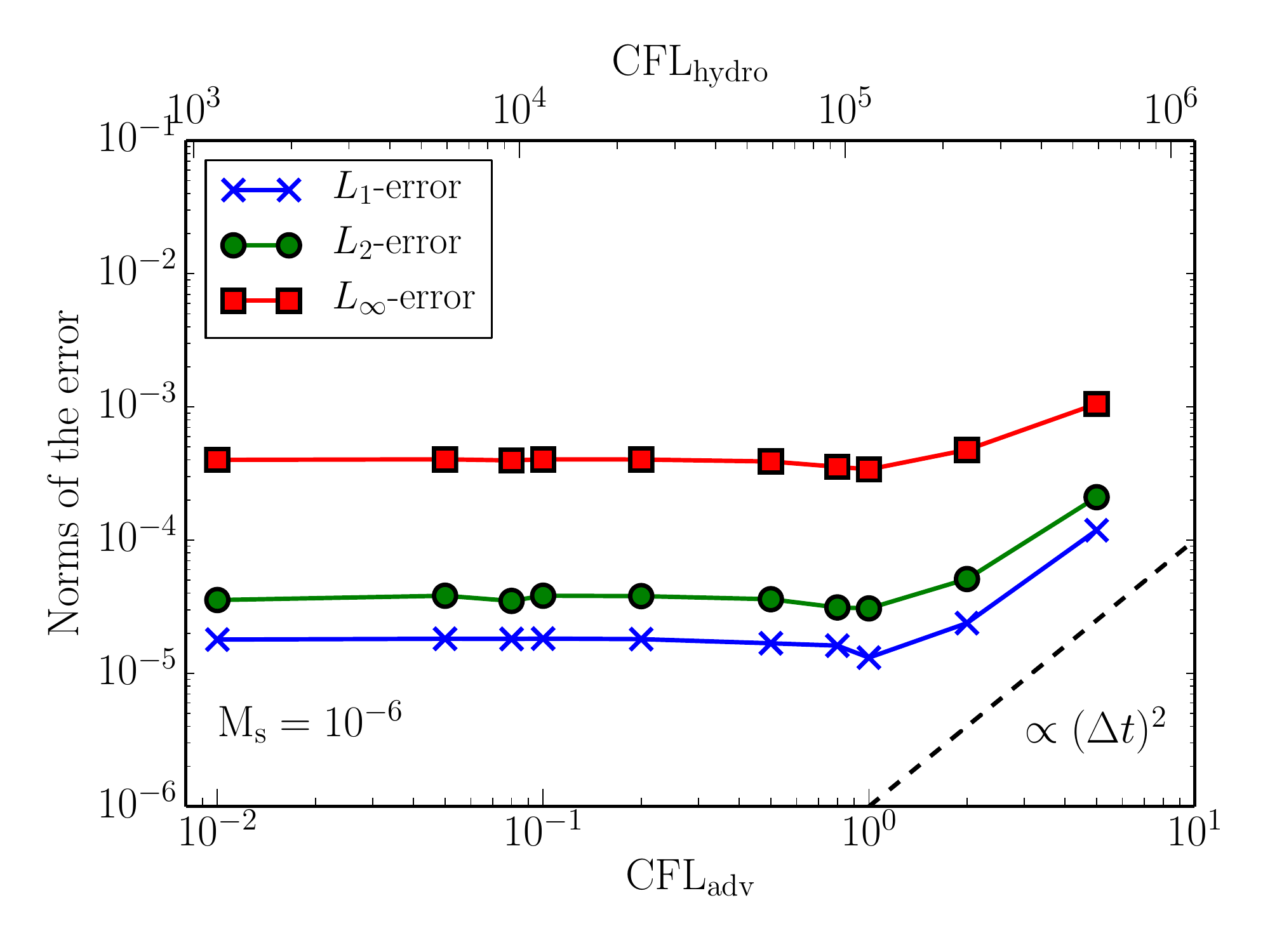}}
   \caption{Same as Fig. \ref{fig:test_si_vortex_adv}, but using the JFNK+PBP scheme to advect the isentropic vortex.}
   \label{fig:full_test_si_vortex_adv}
\end{figure*}

\section{Results}
\label{results}

In this section, we assess the performance of our JFNK+PBP method in both 2D and 3D. In Sect. \ref{ideal_test_cases}, we test the accuracy and efficiency of the method using idealized tests that use an ideal-gas equation of state and a Cartesian geometry. In Sect. \ref{stellar_cases}, we test the method to model stellar interiors in a spherical geometry. An important goal of this section is to demonstrate the good performance, robustness and accuracy of the solver for a wide range of Mach numbers, typically from $M_s=10^{-1}$ down to $M_s=10^{-6}$.


\subsection{Ideal Test Cases}
\label{ideal_test_cases}



\subsubsection{2D isentropic vortex}
\label{jfnk:vortex}

We first investigate the accuracy of the solver by considering the 2D isentropic vortex problem that we used to test the SI scheme in Sect. \ref{si:tests}. We perform the same set of runs with the JFNK+PBP method, and the computed errors are shown in Fig. \ref{fig:full_test_si_vortex_adv}. Comparing with the error of the semi-implicit scheme (see Fig. \ref{fig:test_si_vortex_adv}), the JFNK+PBP scheme achieves an overall reduction in the error. The range of time steps where spatial truncation errors dominate is larger, and we observe the second-order character of the temporal error at large time steps. The use of a SI scheme as a preconditioner does not impact the overall accuracy of the JFNK+PBP method. Figure \ref{fig:full_test_si_vortex_adv} also shows that the results of the JFNK+PBP method are independent of the Mach number, and this desirable property of the SI scheme has been inherited by the nonlinear method.

Four free parameters enter the JFNK+PBP method: the choice of perturbation strategy (Eqs. \ref{pert_old} \& \ref{pert_new}); the two scaling coefficients for the velocity components of the solution and residual vectors (parameters $\alpha_1$ and $\alpha_2$ in Sect. \ref{pbp:nk}); the tolerance required for the solution of the Jacobian equation (parameter $\eta$ in Eq. \ref{eq:forcing_term}). These free parameters were determined by testing.


We find that the accuracy of the solver for the higher end of the Mach number range being considered ($M_s > 10^{-4}$) is good regardless of the choice of perturbation strategy. However, for lower Mach numbers ($M_s \le 10^{-4}$) we find that  only Eq. \eqref{pert_old}, with $\lambda = 10^{-7}$, is able to yield accurate results. When using Eq. \eqref{pert_new}, the Jacobian operator is poorly approximated, regardless of the value of $\lambda$, resulting in a failure of the nonlinear method.

In the scaling of the linear system, we find that a value of $\alpha_1 = 10^{-5}$ gives the most consistent errors across Mach numbers for the range being considered, i.e. $10^{-6} \le M_s \le 10^{-1}$. However, this range can be adjusted by tuning $\alpha_1$ to the problem at hand, with a higher value producing more accurate solutions for high Mach number flows. We find that $\alpha_2 = 1 $ enables us to obtain accurate results for the range of Mach numbers being considered.


We find that a linear tolerance $\eta = 10^{-4}$ produces solutions with similar errors
for the full range of Mach numbers considered in this work.  A choice of $\eta=10^{-2}$ produces similar results for the higher Mach numbers, but the quality of the solutions for low Mach numbers degrades seriously.


Next, we assess the efficiency of the SI scheme as a preconditioner. This is done by considering the number of GMRES iterations necessary to reach convergence without preconditioning and with physics-based preconditioning, for different values of CFL$_\mathrm{hydro}$ and different Mach numbers (the linear tolerance is set to $\eta = 10^{-4}$). When solving the unpreconditioned linear system, we found that for $\alpha_1=10^{-5}$ the majority of linear problems, particularly for higher Mach numbers, fails to converge. Instead, we present for the unpreconditioned case the convergence behavior for $\alpha_1=1$, as a best case scenario. When solving the linear system with physics-based preconditioning, we use the optimal parameters described previously.

\begin{figure*}[t] 
   \centering
      \parbox{0.4\linewidth}{\centering \includegraphics[width=0.8\linewidth, trim= 60 0 0 0]{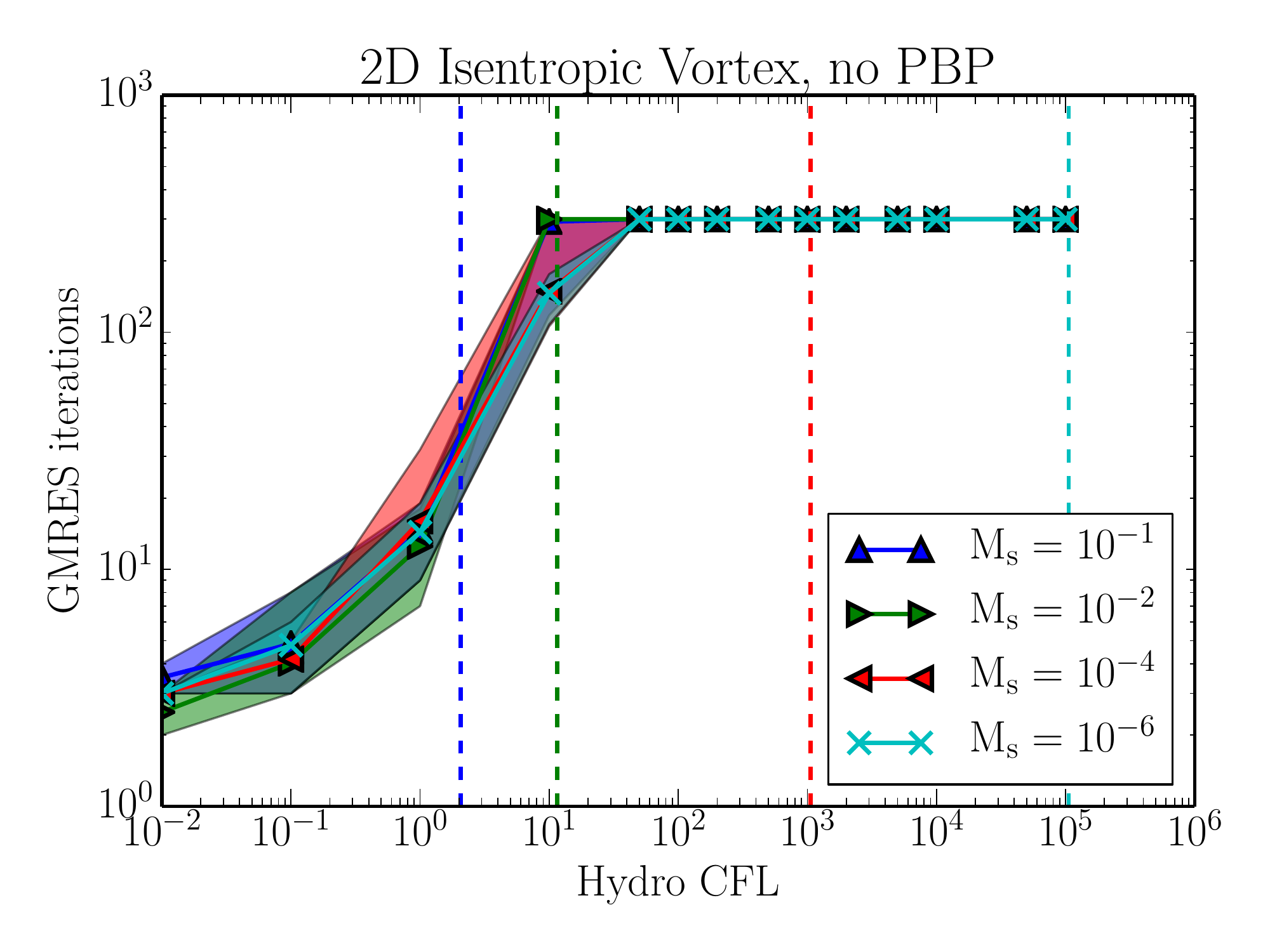}}
      \parbox{0.4\linewidth}{\centering \includegraphics[width=0.8\linewidth, trim= 60 0 0 0]{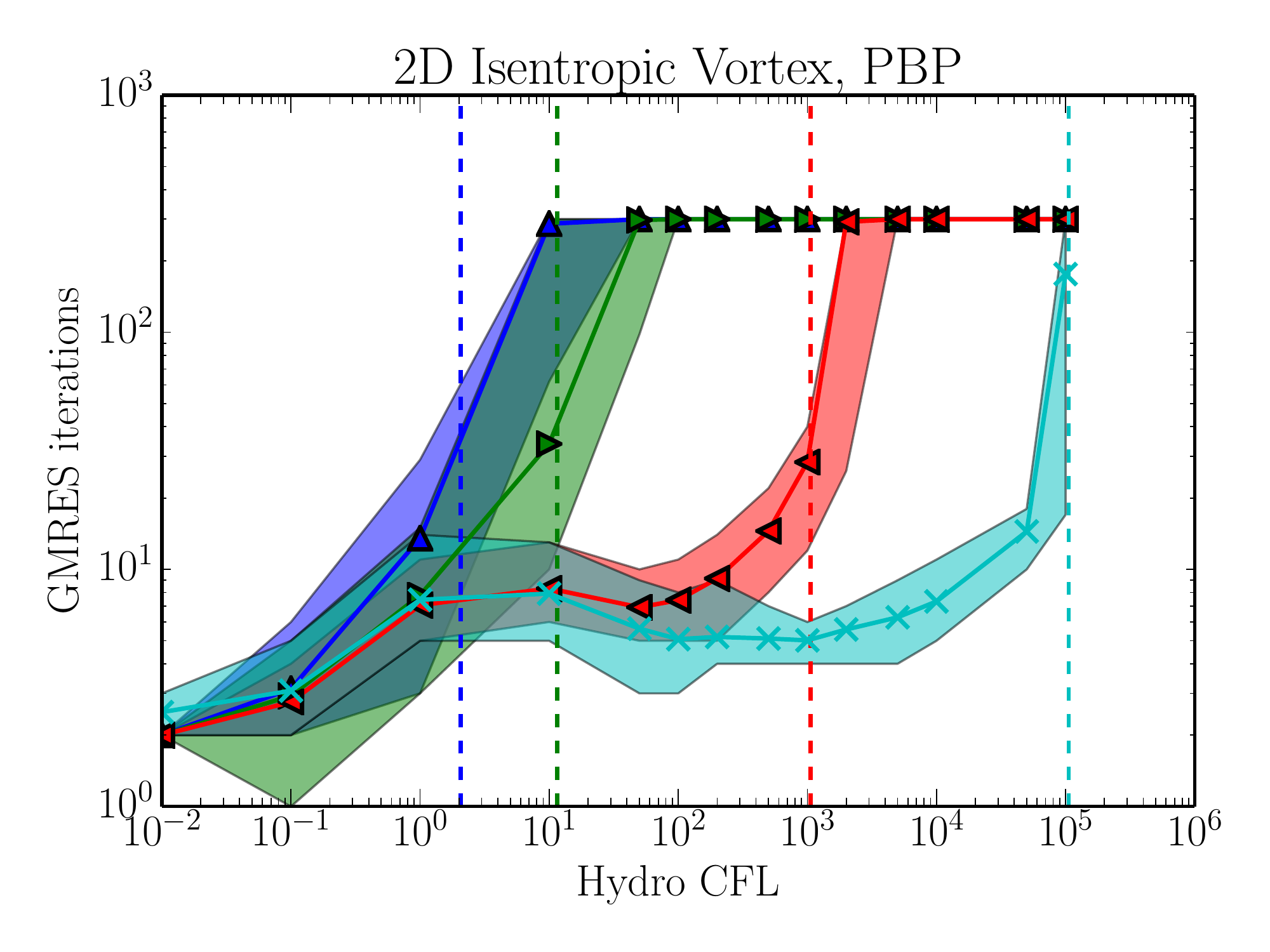}}
   \parbox{0.4\linewidth}{\centering \includegraphics[width=0.8\linewidth, trim= 60 0 0 0]{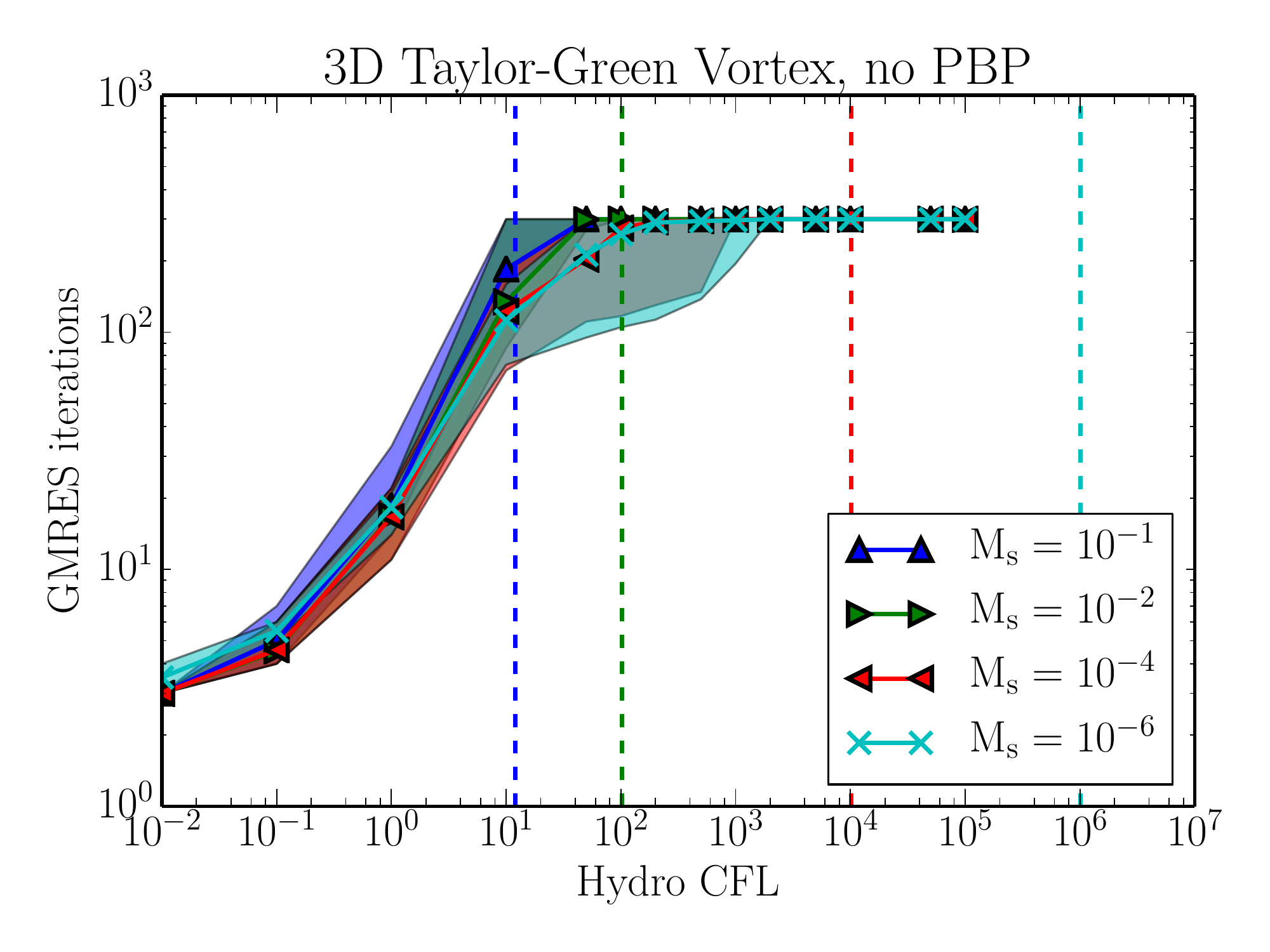}}
      \parbox{0.4\linewidth}{\centering \includegraphics[width=0.8\linewidth, trim= 60 0 0 0]{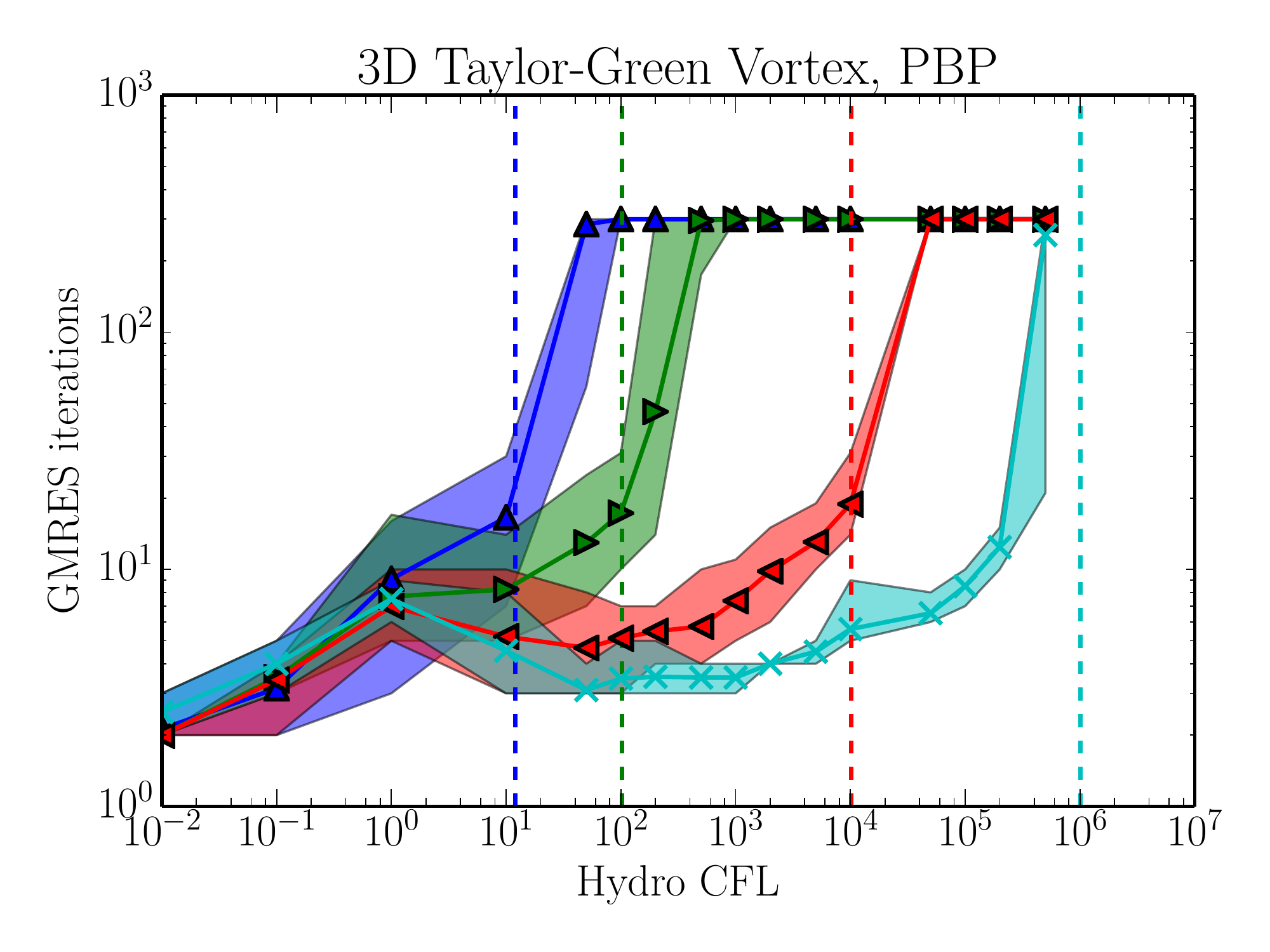}}

   \caption{Convergence of the GMRES solver without preconditioning (left panels) and with the physics-based preconditioner (right panels). The upper panels correspond to the 2D isentropic vortex, and the lower panels to the 3D Taylor-Green vortex. In both cases, the Mach numbers considered are $M_s = 10^{-1}$, $10^{-2}$, $10^{-4}$, $10^{-6}$ (the right panels assume the same legend as the left ones). The maximum allowed number of GMRES iterations was set to 300. The mean values of the number of iterations for convergence is plotted, with shaded areas showing maximum and minimum values. For each Mach number, the location of CFL$_\mathrm{hydro}$ corresponding to CFL$_\mathrm{adv}=1$ is shown by a vertical dashed line.}
   \label{fig:vortex}
   \label{fig:tgv}
\end{figure*}

The results of convergence tests for the iterative method are shown in Fig. \ref{fig:vortex}.  For these tests, the simulations are run for 100 time steps. Without preconditioning, the different Mach number cases behave similarly: the number of GMRES iterations increases rapidly for CFL$_\mathrm{hydro} \gtrsim 1$, and above CFL$_\mathrm{hydro} \gtrsim 10$ no convergence is achieved despite the large number of iterations allowed. Such behavior is due to the stiffness of acoustic waves which increases with CFL$_\mathrm{hydro}$. Our physics-based preconditioner is tailored to treat this effect, and the improvement is demonstrated in the right panel of Fig. \ref{fig:vortex}, as compared to the left panel without preconditioning. In each case, the increase in the number of GMRES iterations takes place at larger values of the time step. With physics-based preconditioning, the failure of the linear solver is now coming from the unstable behavior of the SI scheme for too large a CFL$_\mathrm{adv}$. 

\subsubsection{3D Taylor-Green vortex}

We consider the Taylor-Green vortex problem to test our physics-based preconditioner for an adiabatic (i.e. no thermal diffusion) flow in 3D. We consider a Cartesian domain $(x,y,z) \in [0, 2\pi L]^3$, where $L$ is a lengthscale that sets the physical size of the domain. The initial conditions for the velocity field are

\begin{align}
u_x(x,y,z) &= u_0 \sin \frac{x}{L} \cos \frac{y}{L} \cos \frac{z}{L},\\
u_y(x,y,z) &= - u_0 \cos \frac{x}{L} \sin \frac{y}{L} \cos \frac{z}{L},\\
u_z(x,y,z) & = 0.
\end{align}

\noindent The domain has a uniform density of $\rho_0$. The initial pressure field is

\begin{align}
p(x,y,z) = p_0 + \frac{1}{16} \rho_0 u_0^2 (2 + \cos 2\frac{z}{L})(\cos 2\frac{x}{L} + \cos 2\frac{y}{L}),
\end{align}

\noindent which ensures that

\begin{equation}
\label{eq:tgv_divfree}
\partial_t \big ( \grad \cdot \vec u \big ) = 0\mathrm{\ at\ }t=0,
\end{equation}

\noindent i.e. the initial conditions do not induce any acoustic modes. The initial amplitude of the vortex is measured in terms of the Mach number $M_s = u_0/c_s$, where $c_s = \sqrt{\gamma p_0/\rho_0}$ is the adiabatic sound speed. The adiabatic index $\gamma$ is taken as $7/5=1.4$. We take $L$ as our unit of length, $u_0$ as our unit of velocity, and $\rho_0 L^3$ as our unit of mass. In this normalisation, time is measured in units of $L/u_0$, and energy density in units of $\rho_0 u_0^2$. We change $p_0$ to vary the Mach number in the range $10^{-6} \leq M_s \leq 10^{-1}$. We consider a numerical domain with a resolution of $64^3$. For this test case, we define the advective CFL number as

\begin{equation}
\label{eq:cfl_adv}
\mathrm{CFL}_\mathrm{adv} = \max \frac{ |\vec{u}|  \Delta t}{\Delta x},
\end{equation}


\noindent where $\vec{u}$ is the velocity, $\Delta t$ the time step, $\Delta x$ the mesh spacing.

Similarly to the 2D isentropic vortex, we first investigate the efficiency of the physics-based preconditioner in reducing the number of iterations required by the linear solver to converge to the desired accuracy. As condition (\ref{eq:tgv_divfree}) is never exactly fulfilled in the discretized problem, some acoustic fluctuations are produced at the first time step. To remove these transients, we evolve each case for 100 time steps at a fixed CFL$_\mathrm{hydro}=1$. We then compute another 100 time steps with different values of $\Delta t$, corresponding to different values of CFL$_\mathrm{hydro}$. We monitor the number of iterations required for convergence, with and without physics-based preconditioning\footnote{The parameters of the solver ($\eta$, $\alpha_1$, $\alpha_2$, \dots) are adjusted as discussed for the 2D test case.}. The results are shown in Fig. \ref{fig:tgv}. The conclusions are very similar to the ones drawn from the 2D vortex test presented in the previous section: physics-based preconditioning allows for a fast convergence over a broad range of hydrodynamical CFL numbers. Here again, the convergence of the linear solver becomes difficult when CFL$_\mathrm{adv} =1$ is approached.



\begin{table*}[t]
\caption{Summary of the results for the Taylor-Green vortex tests. The columns for the implicit case represent: the average hydrodynamical and advective CFL numbers, the average number of Newton iterations per time step, the average number of GMRES iterations per Newton iteration, the average number of parabolic iterations for the preconditioner per GMRES iteration.}
\center
\begin{tabular}{c | c c c c c c | c}
\hline
\hline
Mach No. & \multicolumn{6}{c}{Implicit} & Explicit \\ 
\hline
 & CFL$_\mathrm{hydro}$ &  CFL$_\mathrm{adv}$ & $\scriptscriptstyle{\frac{\mathrm{Newton}}{\Delta t}}$ & $\scriptscriptstyle{\frac{\mathrm{GMRES}}{\mathrm{Newton}}}$ & $\scriptscriptstyle{\frac{\mathrm{Parabolic}}{\mathrm{GMRES}}}$ & Final Time & Final Time \\ 
$10^{-1}$ & 7.1e+00 &  0.5 & 3.8 & 16.6 & 2.4 & 17.8 & 34.8\\
$10^{-2}$ & 7.0e+01 &  0.5 & 3.1 & 15.4 & 2.5 & 20.9 & 3.62\\
$10^{-3}$ & 8.2e+02 &  0.5 & 2.7 & 16.2 & 2.5 & 31.6 & 0.300\\
$10^{-4}$ & 1.1e+04 &  0.5 & 2.0 & 16.1 & 2.8 & 55.1 & $3.65(-2)$\\
$10^{-5}$ & 8.1e+04 &  0.5 & 2.0 & 22.2 & 2.9 & 30.9 & $3.64(-3)$\\
$10^{-6}$ & 4.9e+05 &  0.49 & 4.5 & 291.5 & 3.0 & 0.6 & $3.64(-4)$\\
$10^{-6}$ & 5.0e+04 &  0.05 & 2.0 & 7.0 & 2.8 & 4.5 & $3.64(-4)$\\
\end{tabular}
\label{table:tgv2}
\end{table*}

Next we use the Taylor Green vortex to benchmark the implicit JFNK+PBP method against  the second-order accurate Adams-Bashforth explicit scheme. Starting at $t=0$, we evolve the vortex using both the JFNK+PBP method and the Adams-Bashforth method. We run the tests for a fixed wall-clock time of six hours, and record the final time achieved by each method, varying the Mach number of the test case. In the explicit case, the time step is limited by stability to CFL$_\mathrm{hydro} = 0.1$; for the implicit case, the time step is limited to CFL$_\mathrm{adv} = 0.5$ for accuracy and for the stability of the underlying SI. The results are recorded in Table \ref{table:tgv2}. The final times obtained with the explicit solver scale approximately with the Mach number, due to the scaling of the CFL time step with the background sound speed. The final times obtained with the JFNK+PBP method show less of a clear pattern, with performance peaking at a Mach number of $M_s = 10^{-4}$. Nevertheless, the JFNK+PBP method is already more than five times faster than the explicit solver for $M_s=10^{-2}$. For lower Mach numbers, the speed-up is larger than two orders of magnitude. We observe a dramatic drop in performance at $M_s = 10^{-6}$ when using the criterion CFL$_\mathrm{adv}=0.5$ on the time step. From analyzing the performance of the scheme for this run (see Table \ref{table:tgv2}), it appears that the physics-based preconditioner becomes less effective, resulting in a very large number of GMRES iterations and a substantial loss of performance. Such a loss of effectiveness of the preconditioner at a very low-Mach number close to CFL$_\mathrm{adv} \sim 1$ can be already seen on Fig. \ref{fig:tgv}, and seems to highlight the limit of what is currently feasible with the solver. We repeated the timing test for this Mach number with CFL$_\mathrm{Hydro} = 5 \times 10^{4}$, which corresponds to CFL$_\mathrm{adv} \sim 0.05$. This improves the final time by approximately an order of magnitude. It remains the least efficient case, but it is still roughly three orders of magnitude faster than the corresponding explicit calculation.

\begin{figure*}[t] 
  \center
  \includegraphics[width=0.45\linewidth, trim= 0 0 0 0]{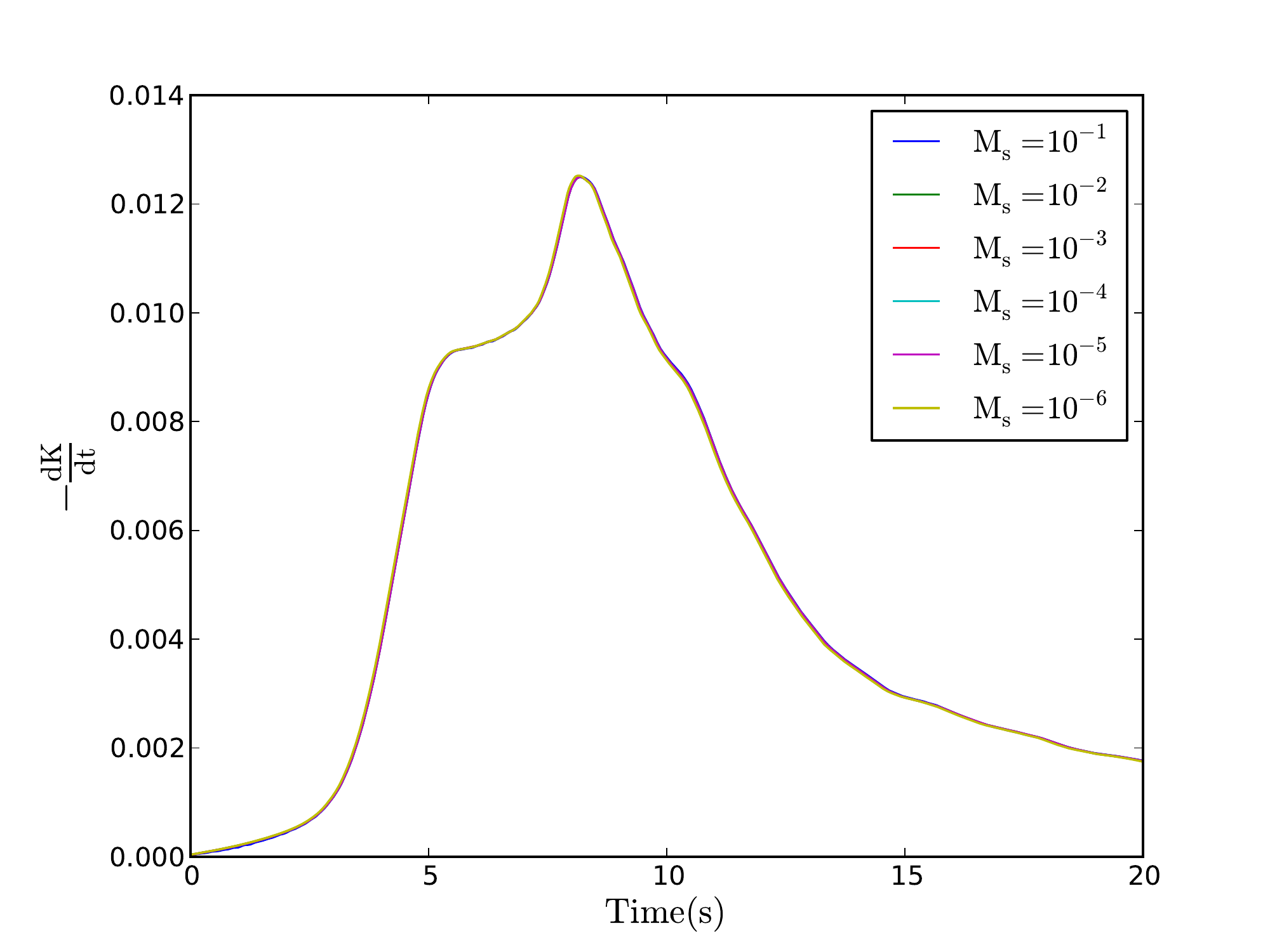}
    \includegraphics[width=0.45\linewidth, trim= 0 0 0 0]{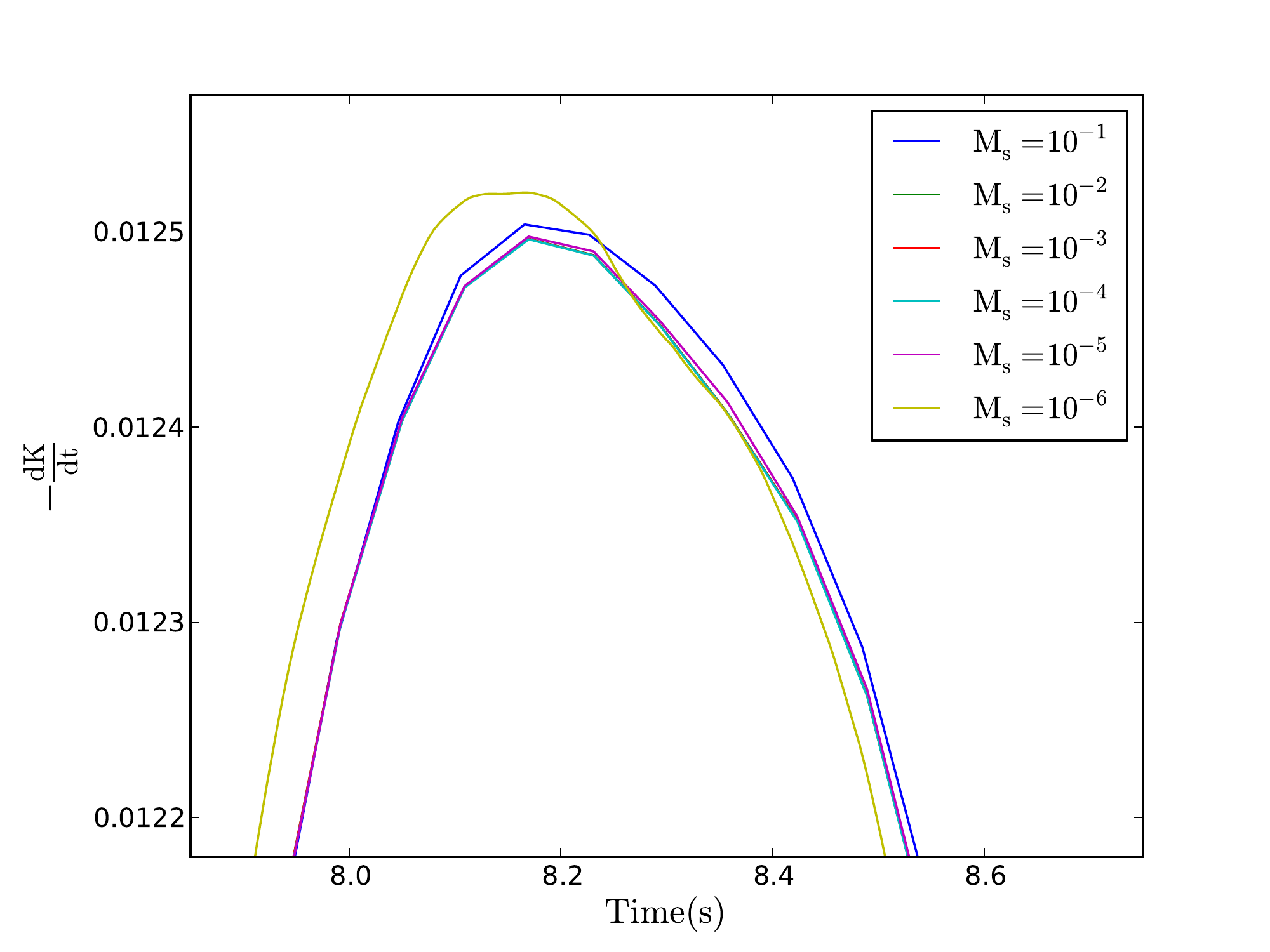}
  \caption{Decay rate of the Taylor Green vortex for different Mach numbers. The right panel shows a zoom on the peak of the decay rate. Time is measured in units of $L/u_0$, the decay rate in units of $\rho_0 u_0^3/L$.}
  \label{fig:tgv_decay}
\end{figure*}

Finally, we monitor the decay of the Taylor-Green vortex for the range of Mach numbers explored here. We simulate for a fixed time of $t = 20$, a time at which most of the dissipation has occurred. We show in Fig. \ref{fig:tgv_decay} the evolution of the decay rate of kinetic energy. The left panel shows a global view where the different curves are indistinguishable from each other. The right panel shows a zoom on the peak of the decay rate. The difference between the curves represents less than a percent. In Table \ref{table:tgv_decay} we record the maximum decay rate and the time at which it occurs. The purpose of Fig. \ref{fig:tgv_decay} is twofold: firstly, it complements the performance results presented previously as it shows that the results are independent, at the percent level, of the Mach numbers; secondly, it provides confidence in using the code as an ILES tool to model turbulent flows over a wide range of Mach numbers. In the ILES framework, dissipation of kinetic energy is due to the truncation errors of the scheme, and it is not obvious that these behave similarly for different Mach numbers.



\subsection{Stellar Test Cases}
\label{stellar_cases}

In this section, we examine how the JFNK+PBP method performs in realistic stellar models. We use the same models as in \cite{2013_comparison} of a 2D young Sun and a red giant in which convection is fully developed and has reached a quasi-steady state. Both models are first considered in a 2D spherically axisymmetric geometry. The red giant model is then considered in a full 3D spherical wedge geometry. 

\begin{table}[t]
\caption{Maximum decay rate measured during the decay of the Taylor Green vortex for different Mach numbers. Time is measured in units of $L/u_0$, the decay rate in units of $\rho_0 u_0^3/L$.}
\center
\begin{tabular}{c|c|c}
  \hline
  \hline
Mach No. & Time of Maximum & Value\\
\hline
$10^{-1}$ & 8.1656 & 1.2504(-2)\\
$10^{-2}$ & 8.1695 & 1.2496(-2)\\
$10^{-3}$ & 8.1695 & 1.2496(-2)\\
$10^{-4}$ & 8.1696 & 1.2496(-2)\\
$10^{-5}$ & 8.1695 & 1.2498(-2)\\
$10^{-6}$ & 8.1681 & 1.2520(-2)\\
\end{tabular}
\label{table:tgv_decay}
\end{table}

\subsubsection{2D stellar models}
\label{2Dstellar_tests}

\begin{table*}[t]
\caption{Comparison of the performance of the JFNK+PBP method presented in this paper with the Broyden methods presented in \cite{2013_comparison}, for the 2D red giant test case. The value of the linear tolerance $\eta$ is given in parenthesis after the name of the method; ILU($k$) refers to an incomplete LU factorization of order $k$; PBP refers to physics-based preconditioning for sound waves only.}
\label{tab:rg_runs_summary}
\centering

\begin{tabular}{l c c c c c c}
\hline \hline
Method & CFL$_\mathrm{hydro}$ &  CFL$_\mathrm{adv}$ & CFL$_\mathrm{rad}$ & $\scriptscriptstyle{\frac{\mathrm{Newton}}{\Delta t}}$ & $\scriptscriptstyle{\frac{\mathrm{GMRES}}{\mathrm{Newton}}}$ & $\scriptscriptstyle{\frac{\mathrm{Simulated\ time}}{\mathrm{Wall\ time}}}$ \\
\hline
{\bf CFL$_\mathrm{adv,max} = 0.5$} \\
Broyden($10^{-2}$)+ILU(1) & 17.7 & 0.46 & 4.8 & 7.3 & 4.0 & 227\\
JFNK($10^{-1}$)+ PBP & 18.9 & 0.49 & 4.9 & 6.2 & 7.2 & 202\\
JFNK($10^{-2}$)+ PBP & 18.9 & 0.49 & 4.9 & 4.6 & 18.4 & 161\\
JFNK($10^{-4}$)+PBP & 18.9 & 0.49 & 4.9 & 4.6 & 37.8 & 80\\

\hline
{\bf CFL$_\mathrm{adv,max} = 1$} \\
Broyden($10^{-1}$)+ILU(1) & 37.9 & 0.93 & 8.2 & 9.5 & 4.3 & 382\\
JFNK($10^{-1}$)+PBP & 40.2 & 0.98 & 8.6 & 6.6 & 18.5 & 200\\
JFNK($10^{-2}$)+PBP & 40.0 & 0.98 & 8.5 & 5.7 & 41.3 & 124\\
JFNK($10^{-4}$)+PBP & 40.0 & 0.98 & 8.5 & 5.7 & 200.7 & 32\\
\hline
{\bf CFL$_\mathrm{adv,max} = 1.5$} \\
Broyden($10^{-1}$)+ILU(1) & 45.2 & 1.10 & 9.6 & 13.2 & 4.9 & 383\\
JFNK($10^{-1}$)+PBP & 55.2 & 1.42 & 11.0 & 8.9 & 37.8 & 121\\
JFNK($10^{-2}$)+PBP & 55.8 & 1.43 & 11.0 & 7.8 & 79.3 & 73\\
JFNK($10^{-4}$)+PBP & 55.7 & 1.43 & 11.0 & 7.7 & 252.0 & 33\\
\end{tabular}
\end{table*}

\begin{table*}[t]
   \caption{Similar to Table \ref{tab:rg_runs_summary}, for the 2D young Sun models.}
   \label{tab:ys_runs_summary}
   \centering

\begin{tabular}{l c c c c c c}
\hline \hline
Method & CFL$_\mathrm{hydro}$ &  CFL$_\mathrm{adv}$ & CFL$_\mathrm{rad}$ & $\frac{\mathrm{Newton}}{\Delta t}$ & $\frac{\mathrm{GMRES}}{\mathrm{Newton}}$ & $\frac{\mathrm{Simulated\ time}}{\mathrm{Wall\ time}}$ \\
\hline
{\bf CFL$_\mathrm{adv,max} = 0.5$} \\
Broyden($10^{-1}$)+ILU(2) & 235.5 & 0.50 & $2.5(-7)$ & 6.6 & 12.3 & 74\\

JFNK($10^{-1}$)+PBP & 235.0 & 0.50 & $2.5(-7)$ & 6.7 & 10.4 & 39\\
JFNK($10^{-2}$)+PBP & 227.4 & 0.50 & $2.4(-7)$ & 5.2 & 15.8 & 35\\
JFNK($10^{-4}$)+PBP & 235.0 & 0.50 & $2.5(-7)$ & 5.2 & 19.9 & 28\\


\hline
{\bf CFL$_\mathrm{adv,max} = 1$} \\
Broyden($10^{-1}$)+ILU(2) & 474.8 & 1.00 & 5.1(-7) & 8.2 & 16.7 & 106\\
JFNK($10^{-1}$)+PBP & 474.9 & 1.00 & 5.1(-7) & 7.5 & 19.5 & 39\\
JFNK($10^{-2}$)+PBP & 471.8 & 0.99 & 5.1(-7) & 8.1 & 20.0 & 35\\
JFNK($10^{-4}$)+PBP & 471.8 & 0.99 & 5.1(-7) & 8.0 & 20.0 & 34\\
\hline
{\bf CFL$_\mathrm{adv,max} = 1.5$} \\
Broyden($10^{-2}$)+ILU(2) & 681.3 & 1.50 & 7.3(-7) & 9.9 & 19.8 & 121\\
JFNK($10^{-1}$)+PBP & 662.8 & 1.45 & 7.1(-7) & 15.6 & 19.9 & 26\\
JFNK($10^{-2}$)+PBP & 656.9 & 1.44 & 7.1(-7) & 15.9 & 20.0 & 25\\
JFNK($10^{-4}$)+PBP & 668.0 & 1.46 & 7.2(-7) & 16.0 & 20.0 & 25\\
\end{tabular}
\end{table*}

We compute 100 time steps of the red giant and young Sun models using the JFNK+PBP method. We limit CFL$_\mathrm{adv}$ (defined as in Eq. \ref{eq:cfl_adv}) to values of 0.5, 1, and 1.5. We compare the performance of the JFNK+PBP method with the best method identified in \cite{2013_comparison}. The results are summarized in Table \ref{tab:rg_runs_summary} for the red giant, and Table \ref{tab:ys_runs_summary} for the young Sun. They show that the JFNK+PBP method is less efficient than the Broyden+ILU method. It is seen that the JFNK+PBP method is becoming less and less effective when CFL$_\mathrm{adv}$ increases, as the physics-based preconditioner fails as the underlying SI becomes unstable. In practice, the JFNK+PBP method should not be used with CFL$_\mathrm{adv}$ larger than one when computing an unsteady flow. This limitation is not very penalizing, as numerical accuracy is expected to decrease when CFL$_\mathrm{adv} > 1$, meaning that larger time steps are not desirable anyway\footnote{Based on the 2D vortex advection test, \cite{2013_comparison} showed that one could use CFL$_\mathrm{adv} \sim 2$ without degrading the accuracy too much. However, this conclusion might not be adequate for an unsteady, turbulent flow, which could require smaller time step.}. For both the red giant and young Sun models, the performance of the JFNK+PBP solver is the same for CFL$_\mathrm{adv}=0.5$ and CFL$_\mathrm{adv}=1$, as the increase in the time step is compensated by the increase in the number of GMRES iterations per Newton iteration. One should keep in mind that the performance of the JFNK+PBP solver presented here could probably be improved by fine tuning the parameters discussed in Sect. \ref{jfnk:vortex}. For instance, the red giant and young Sun models differ in the average Mach number, with the red giant having a larger Mach number ($\sim 0.1$) than the young Sun ($\sim 0.01$). Although the performance of the JFNK+PBP method could be made closer to the Broyden method, we could expect the latter to remain the most efficient option for these cases.

\subsubsection{3D red giant models}

\begin{table*}[t]
\caption{Similar to Table \ref{tab:rg_runs_summary}, for the 3D red giant models.}
   \label{tab:rg_runs_summary_3D}
   \centering
\begin{tabular}{l c c c c c c}
\hline \hline
Method & CFL$_\mathrm{hydro}$ &  CFL$_\mathrm{adv}$ & CFL$_\mathrm{rad}$ & $\frac{\mathrm{Newton}}{\Delta t}$ & $\frac{\mathrm{GMRES}}{\mathrm{Newton}}$ & $\frac{\mathrm{Simulated\ time}}{\mathrm{Wall\ time}}$ \\
\hline
{\bf CFL$_\mathrm{adv,max} = 0.5$} \\
JFNK($10^{-1}$)+PBP & 107.6 & 0.50 & 0.74 & 10.8 & 7.8 & 188\\
JFNK($10^{-2}$)+PBP & 107.6 & 0.50 & 0.74 & 7.3 & 13.3 & 183\\
JFNK($10^{-4}$)+PBP & 107.6 & 0.50 & 0.74 & 5.1 & 25.6 & 147\\
Broyden($10^{-1}$) w/o preconditioner& 107.6 & 0.50 & 0.74 & 7.6 & 219.7 & 122\\
Broyden($10^{-1}$)+ILU(2) & 107.6 & 0.50 & 0.74 & 6.8 & 8.7 &109\\
\hline
{\bf CFL$_\mathrm{adv,max} = 1$} \\
Broyden($10^{-1}$)+ILU(2) & 234.5 & 1.00 & 1.4 & 8.7 & 14.0 & 203\\
Broyden($10^{-1}$) w/o preconditioner & 234.6 & 1.00 & 1.4 & 10.4 & 268.1 & 180\\
JFNK($10^{-1}$)+PBP & 227.4 & 0.97 & 1.4 & 21.5 & 15.7 &  109\\
\hline
{\bf CFL$_\mathrm{adv,max} = 1.5$} \\
Broyden($10^{-1}$)+ILU(2) & 342.5 & 1.50 & 1.9 & 10.9 & 17.1 &257\\
Broyden($10^{-1}$) w/o preconditioner & 342.5 & 1.50 & 1.9 & 12.7 & 286.0 &205\\
JFNK($10^{-1}$)+PBP & 227.0 & 0.97 & 1.4 & 25.0 & 22.5 & 67\\
\end{tabular}
\end{table*}

The efficient computation of 3D models is the main motivation for moving beyond the framework of quasi-Newton methods. We cannot, however, meaningfully compare the performance of the later method to that of the JFNK+PBP method for 3D stellar models, as done in the previous section. As shown previously, quasi-Newton methods, such as the Broyden method, perform well in 2D. However, their cost increases significantly in 3D. The reasons are twofold. Firstly, in 3D, the Jacobian matrix has a more complex structure than in 2D, due to the third dimension.  This implies an increase in the cost for the construction and storage of the Jacobian matrix and its ILU factorization. As a result, for the same number of degrees of freedom (i.e. same matrix size), a 3D computation is inherently more expensive than a 2D computation. Secondly, in 3D, the typical size of a problem is much larger than in 2D, essentially due to the larger number of cells, but also due to the extra variable (the third velocity component). For instance, a $128^2$ computation has  $4\times128^2 =65536$ degrees of freedom, whereas a $128^3$ computation has $5\times128^3= 10\ 485\ 760$ degrees of freedom. The computational costs (cpu time+memory) for some of the components of the quasi-Newton methods do not scale linearly with the problem size. Thus, this increase in degrees of freedom corresponds to a prohibitive increase in both cpu time and memory.
For these reasons, we can only perform a comparison with an extremely low resolution, not necessarily relevant to the analysis of physical processes in stars. Since the JFNK+PBP is now the method implemented in \MUSIC\ for the purpose of running 3D simulations, we want to illustrate the potential of this method.

We do so by performing computations of the red giant model for a grid size of $72\times65^2$ (roughly 1.5 million degrees of freedom), using both the Broyden and JFNK+PBP methods. For the reasons presented previously, this is the largest problem size that we could consider using the serial version of \MUSIC\ on a single node of the supercomputer Zen at the University of Exeter. Each node has 12 cores and 24 Gb of RAM, and a full node was requested to benefit from the available memory. It is clear that the memory requirement of the ILU factorization restricts the range of accessible resolutions, even if domain decomposition is used to distribute the problem among several computer nodes.

Test runs are performed similarly to the previous section, and the results are summarized in Table \ref{tab:rg_runs_summary_3D}. The JFNK+PBP method is more efficient for CFL$_\mathrm{adv}=0.5$, but the Broyden method remains more efficient for CFL$_\mathrm{adv}=1$ and CFL$_\mathrm{adv}=1.5$. Surprisingly enough, for this particular case the unpreconditioned Broyden method performs almost as well as the preconditioned version, showing that the ILU preconditioner becomes inefficient due to its cost. However, we stress that an unpreconditioned Broyden method is not a viable option for scientific applications. The preconditioned version will not be viable for larger problems, due to the increasing cost for computing and storing the ILU factorization. We expect a more substantial gain compared to the quasi-Newton methods when larger problems will be considered, but the serial tests performed here limit us to relatively small 3D problems. Such small problems appear already to be on the edge of the capabilities of quasi-Newton methods. The implementation of the JFNK+PBP method in \MUSIC\ now allows us to perform 3D simulations with resolutions of $512^3$ of a large fraction ($\sim 80$\% in radius) of a partly convective star, as an initial step toward the study of turbulent convection and overshooting under realistic stellar interior conditions and over relevant physical timescales (Pratt et al., in prep.)

Finally, as stellar models include radiative diffusion, we have the possibility of using the second version of the physics-based preconditioner, in which thermal diffusion is also treated implicitly. The stiffness of thermal diffusion is measured by the radiative CFL number, defined as:

\begin{equation}
\mathrm{CFL}_\mathrm{rad} = \max \frac{\chi  \Delta t}{\Delta x^2},
\end{equation}
\noindent where $\Delta t$ is the time step, $\Delta x$ the mesh spacing, $\chi$ the thermal diffusivity. As for sound waves, solving thermal diffusion with a time explicit method requires CFL$_\mathrm{rad} \lesssim 1$ for stability. Implicit methods allow for CFL$_\mathrm{rad} \gg 1$, but preconditioning is necessary to improve the convergence of the iterative method. For the red giant model, however, our particular treatment of the surface implies that the radiative diffusion is not very stiff (CFL$_\mathrm{rad} \sim 1$), and as such we do not see a substantial difference between the two versions of the physics-based preconditioner. Concrete examples of stellar cases where preconditioning of thermal diffusion is necessary will be presented elsewhere.

%
%

\section{Conclusion}
\label{conclusion}

This work is a continuation of previous efforts devoted to the development of an efficient, accurate fully implicit solver for multidimensional hydrodynamics. In Sect. \ref{pbp} we presented a Jacobian-free Newton-Krylov method, which avoids the explicit construction of the Jacobian matrix. The use of iterative methods to solve the Jacobian equation requires preconditioning at large hydrodynamical CFL numbers. The main purpose of this paper was to present an efficient preconditioner that specifically targets the physical processes that are responsible for numerical stiffness, hence the name of ``physics-based'' preconditioners. This strategy is very different from the more usual algebraic preconditioners (as, e.g., ILU factorization) which try to address the stiffness of the system by looking at the Jacobian matrix structure and numerical values only, without considering the underlying equations. In the context of stellar hydrodynamics, stiffness results from acoustic perturbations that propagate on a time scale much shorter than the fluid bulk motion, and possibly from thermal diffusion. Therefore, the preconditioning step relies on a semi-implicit solver, which is inexpensive and rather inaccurate\footnote{We applied Picard linearization to derive the scheme and used first-order time discretization methods.}, that treats sounds waves (and thermal diffusion, if required) implicitly in order to overcome the associated CFL limit on the time step (see Sect. \ref{SI}). Although we aim at using \MUSIC\ to model stellar interiors, the JFNK+PBP method can be applied to general advection and/or diffusion dominated problems. Although many approximations enter the derivation of our SI scheme, they do not restrict its range of applicability.

Section \ref{results} presented the results of extensive tests assessing the performance and accuracy of the new method. A strong emphasis was put on exploring a wide range of Mach numbers, namely six orders of magnitude from $M_s=10^{-1}$ down to $M_s=10^{-6}$. The tests assessed the ability of the physics-based preconditioner to reduce the number of linear iterations required by the linear solver. Using the 3D Taylor-Green vortex test, we showed that this solver is computationally efficient, beating the Adams-Bashforth explicit scheme for $M_s \lesssim 10^{-2}$. We emphasize that the method has several parameters that can be tuned to improve its performance. In order to achieve the best performance, these parameters should be tuned for the particular problem being considered. Therefore, we do not claim to have found the best set of parameters, but rather a set that gives very satisfying performance for the various tests performed in this paper. Furthermore, the performance does not come with any loss of accuracy: our method exhibits accuracy consistent with the second-order nature of its discretization, and most importantly, the numerical errors are independent of the Mach number, at least in the investigated range. However, it appears that $M_s \sim 10^{-6}$ is on the edge of the abilities of the solver, as fine-tuning of some parameters and a reduction of the time step was necessary to pass the tests at such a low Mach number.

The JFNK+PBP method is now the work-horse of the \MUSIC\ code, and is used to investigate long-standing problems in stellar hydrodynamics such as shear mixing and convective overshooting. Further developments are now devoted to the parallelization of the method, in order to take advantage of multi-cores/multi-nodes high-performance computers that are now routinely used in computational physics. Performance and scalability tests will be presented elsewhere.

\begin{acknowledgement}
MV would like to thank Dana Knoll and Ryosuke Park for useful discussions on physics-based preconditioning during several visits at LANL that motivated this work. MV also thanks Hannes Grimm-Strele and Philipp Edelmann for discussions related to the work presented in this paper. MV acknowledges support  from a Newton International Fellowship and Alumni program from the Royal Society during earlier part of this work. Part of this work was funded by the Royal Society Wolfson Merit award WM090065, the Consolidated STFC grant  ST/J001627/1STFC, the French ``Programme National de Physique Stellaire'' (PNPS) and ``Programme National Hautes Energies'' (PNHE), and by the European Research Council through grants ERC-AdG No. 320478-TOFU and ERC-AdG No. 341157-COCO2CASA. The calculations for this paper were performed on the DiRAC Complexity machine, jointly funded by STFC and the Large Facilities Capital Fund of BIS, and the University of Exeter Supercomputer, a DiRAC Facility jointly funded by STFC, the Large Facilities Capital Fund of BIS, and the University of Exeter.
\end{acknowledgement}

\bibliographystyle{aa}
\bibliography{references}

\appendix

\section{Evolution equation for pressure and acoustic fluctuations}

\subsection{Evolution equation for pressure}
\label{appendix:pressure_equation}

The linearized equation-of-state yields

\begin{equation}
\label{eq:linEOS}
\delta p = \frac{\partial p}{\partial \rho} \Big |_e \delta \rho + \frac{\partial p}{\partial e} \Big |_\rho \delta e.
\end{equation}

\noindent We have the thermodynamic relationships

\begin{align}
\frac{\partial p}{\partial \rho} \Big |_e = & \frac{p}{\rho} \big (\Gamma_1 - \Gamma_3 +1 \big ),\\
\frac{\partial p}{\partial e} \Big |_\rho = &\rho \big ( \Gamma_3 -1 \big ),
\end{align}

\noindent where $\Gamma_1$ and $\Gamma_3$ are the first and third generalised adiabatic indices. We now substitute $\delta$'s with Lagrangian derivatives $D_t = \partial_t + \vec u \cdot \grad$ in Eq. (\ref{eq:linEOS}):

\begin{equation}
D_t p = \frac{p}{\rho} \big (\Gamma_1 - \Gamma_3 +1 \big ) D_t \rho + \rho \big ( \Gamma_3 -1 \big ) D_t e,
\end{equation}

\noindent and we use the Lagrangian equations

\begin{align}
D_t \rho =& - \rho \grad \cdot \vec u,\\
\rho D_t e = & - p \grad \cdot \vec u + \grad \cdot \big ( \chi \grad T \big ),
\end{align}

\noindent to obtain

 \begin{equation}
 \label{eq:pressure}
 \partial_t p + \vec u \cdot \vec \nabla p = - \Gamma_1 p \vec \nabla \cdot \vec u + (\Gamma_3-1) \vec \nabla \cdot \big ( \chi \vec \nabla T \big ).
 \end{equation}

\section{Transformation matrices}
\label{appendix:transformation}

\subsection{Internal energy equation}

In this case $U=(\rho, \rho e, \rho u)$, $X = (\rho, e, u)$, and $V=(p,e,u)$. The transformation matrix $\partial U / \partial X$ between variables $U$ and $X$ is such that $\delta U = (\partial U / \partial X) \delta X$. We have

\begin{equation}
\frac{\partial U}{\partial X} =
\begin{pmatrix}
1 & 0        & 0\\
e & \rho    & 0\\
u & 0        &  \rho \\
\end{pmatrix}.
\end{equation}

\noindent The inverse transformation is:

\begin{equation}
\frac{\partial X}{\partial U} = \Big( \frac{\partial U}{\partial X} \Big)^{-1}
=
\begin{pmatrix}
1 & 0        & 0\\
-e/\rho & 1/\rho & 0\\
-u/\rho &  0 & 1/\rho\\
\end{pmatrix}.
\end{equation}

\noindent The transformation matrix $\partial V / \partial X$ is

\begin{equation}
\frac{\partial V}{\partial X} =
\begin{pmatrix}
 \frac{\partial p}{\partial \rho} \big |_e & \frac{\partial p}{\partial e} \big |_\rho   & 0\\
0 & 1      & 0\\
0 & 0       &  1 \\
\end{pmatrix},
\end{equation}

\noindent and its inverse $\partial X / \partial V$ is

\begin{equation}
\frac{\partial X}{\partial V} =
\begin{pmatrix}
\big ( \frac{\partial p}{\partial \rho} \big |_e \big )^{-1} & - \big ( \frac{\partial p}{\partial \rho} \big |_e \big )^{-1} \frac{\partial p}{\partial e} \Big |_\rho   & 0\\
0 & 1      & 0\\
0 & 0       &  1 \\
\end{pmatrix}.
\end{equation}

\noindent We have

\begin{align}
\frac{\partial V}{\partial U} =& \frac{\partial V}{\partial X} \times \frac{\partial X}{\partial U} \notag \\
 = &
\begin{pmatrix}
 \frac{\partial p}{\partial \rho} \big |_e & \frac{\partial p}{\partial e} \big |_\rho   & 0\\
0 & 1      & 0\\
0 & 0       &  1 \\
\end{pmatrix}
\begin{pmatrix}
1 & 0        & 0\\
-e/\rho & 1/\rho & 0\\
-u/\rho &  0 & 1/\rho\\
\end{pmatrix} \notag \\
 =&
\begin{pmatrix}
\frac{\partial p}{\partial \rho} \big |_e - \frac{e}{\rho} \frac{\partial p}{\partial e} \Big |_\rho  & \frac{1}{\rho} \frac{\partial p}{\partial e} \Big |_\rho  & 0\\
-e/\rho & 1/\rho & 0\\
-u/\rho &  0 & 1/\rho\\
\end{pmatrix}.
\end{align}

\subsection{Total energy equation}

In this case $U=(\rho, \rho \epsilon_t, \rho u)$, $X = (\rho, \epsilon_t, u)$, and $V=(p,e,u)$. The transformation matrix $\partial U / \partial X$ is:

\begin{equation}
\frac{\partial U}{\partial X} =
\begin{pmatrix}
1 & 0        & 0\\
\epsilon_t &  \rho & 0 \\
u &  0  & \rho \\
\end{pmatrix}.
\end{equation}

\noindent Its inverse is

\begin{equation}
\frac{\partial X}{\partial U} =
\begin{pmatrix}
1 & 0        & 0\\
-\epsilon_t /\rho &  1/\rho & 0 \\
-u/\rho & 0  & 1/\rho\\
\end{pmatrix}.
\end{equation}

%

\noindent The transformation matrix from $(\rho, e, u)$ and $(\rho, \epsilon_t, u)$ is

 \begin{equation}
\begin{pmatrix}
\delta \rho \\
\delta \epsilon_t\\
\delta u\\
\end{pmatrix}
=
\begin{pmatrix}
1 & 0   & 0\\
0 &  1 & u \\
0 &  0  & 1 \\
\end{pmatrix}
\begin{pmatrix}
\delta \rho \\
\delta e\\
\delta u\\
\end{pmatrix},
\end{equation}

\noindent so that the transformation matrix $\partial V / \partial X$ is

\begin{align}
\frac{\partial V}{\partial X} =&
\begin{pmatrix}
 \frac{\partial p}{\partial \rho} \big |_e & \frac{\partial p}{\partial e} \big |_\rho   & 0\\
0 & 1      & 0\\
0 & 0       &  1 \\
\end{pmatrix}
\begin{pmatrix}
1 & 0   & 0\\
0 &  1 & u \\
0 &  0  & 1 \\
\end{pmatrix}^{-1} \notag \\
=&
\begin{pmatrix}
 \frac{\partial p}{\partial \rho} \big |_e & \frac{\partial p}{\partial e} \big |_\rho   & 0\\
0 & 1      & 0\\
0 & 0       &  1 \\
\end{pmatrix}
\begin{pmatrix}
1 & 0   & 0\\
0 &  1 & -u \\
0 &  0  & 1 \\
\end{pmatrix} \notag \\
=&
\begin{pmatrix}
 \frac{\partial p}{\partial \rho} \big |_e & \frac{\partial p}{\partial e} \big |_\rho   & -u  \big ( \frac{\partial p}{\partial e} \big |_\rho \big )\\
0 & 1    & -u\\
0 & 0    & 1 \\
\end{pmatrix}.
\end{align}

\noindent The inverse transformation is

\begin{align}
\frac{\partial X}{\partial V} =&
\begin{pmatrix}
1 & 0   & 0\\
0 &  1 & u \\
0 &  0  & 1 \\
\end{pmatrix}
\begin{pmatrix}
 \frac{\partial p}{\partial \rho} \big |_e & \frac{\partial p}{\partial e} \big |_\rho   & 0\\
0 & 1      & 0\\
0 & 0       &  1 \\
\end{pmatrix}^{-1} \notag \\
=&
\begin{pmatrix}
1 & 0   & 0\\
0 &  1 & u \\
0 &  0  & 1 \\
\end{pmatrix}
\begin{pmatrix}
\big ( \frac{\partial p}{\partial \rho} \big |_e \big )^{-1} & - \big ( \frac{\partial p}{\partial \rho} \big |_e \big )^{-1} \frac{\partial p}{\partial e} \Big |_\rho   & 0\\
0 & 1      & 0\\
0 & 0       &  1 \\
\end{pmatrix} \notag \\
=&
\begin{pmatrix}
\big ( \frac{\partial p}{\partial \rho} \big |_e \big )^{-1} & - \big ( \frac{\partial p}{\partial \rho} \big |_e \big )^{-1} \frac{\partial p}{\partial e} \Big |_\rho   & 0\\
0 & 1 & u\\
0 & 0 & 1\\
\end{pmatrix}.
\end{align}

\noindent Finally, we have

\begin{align}
\frac{\partial V}{\partial U} =& \frac{\partial V}{\partial X} \times \frac{\partial X}{\partial U} \notag \\
 = &
\begin{pmatrix}
 \frac{\partial p}{\partial \rho} \big |_e & \frac{\partial p}{\partial e} \big |_\rho   & -u  \big ( \frac{\partial p}{\partial e} \big |_\rho \big )\\
0 & 1    & -u\\
0 & 0    & 1 \\
\end{pmatrix}
\begin{pmatrix}
1 & 0        & 0\\
-\epsilon_t / \rho & 1/\rho & 0\\
-u/\rho &  0 & 1/\rho\\
\end{pmatrix} \notag \\
 =&
\begin{pmatrix}
\frac{\partial p}{\partial \rho} \big |_e - \frac{\epsilon_t-u^2}{\rho} \frac{\partial p}{\partial e} \Big |_\rho  & \frac{1}{\rho} \frac{\partial p}{\partial e} \Big |_\rho  &  - \frac{u}{\rho} \frac{\partial p}{\partial e} \Big |_\rho\\
- (\epsilon_t-u^2)/\rho & 1/\rho & -u/\rho\\
-u/\rho &  0 & 1/\rho\\
\end{pmatrix}.
\end{align}

\end{document}